\documentclass[manuscript,screen]{acmart}
\pdfoutput=1
\usepackage{soul}
\usepackage{booktabs} 
\usepackage{multicol}
\usepackage{multirow}
\usepackage{amsmath,amssymb,amsfonts}
\usepackage{algorithm}
\usepackage{algpseudocode}
\usepackage{subfig}
\usepackage{listings}
\usepackage{graphicx}
\usepackage{tikz}
\usepackage{pgfplots}
\usepackage{textcomp}
\usepackage{svg}
\usepackage{tabularx}
\usepackage{float}
\usepackage{siunitx}

\expandafter\newcommand\csname r@tocindent4\endcsname{4in}

\newcommand\blfootnote[1]{%
  \begingroup
  \renewcommand\thefootnote{}\footnote{#1}%
  \addtocounter{footnote}{-1}%
  \endgroup
}

\setcitestyle{super,sort&compress}
\begin{document}
\title[Optimizing the LiFE Algorithm for Multi-Core and Many-Core Systems]{Optimizing the Linear Fascicle Evaluation Algorithm for Multi-Core and Many-Core Systems}

\titlenote{\textit{Extension of Conference Paper} \\Optimizing the Linear Fascicle Evaluation Algorithm for Many-Core Systems, Karan Aggarwal, Uday Bondhugula, ACM International Conference on Supercomputing (ICS), Jun 2019, Arizona, USA.\\ We made following novel contributions in this work. First, we generalized the data restructuring methods and computation splitting techniques, these are extended to CPUs. Second, we present CPU-specific optimizations to improve the performance of the LiFE applications. Third, we describe DSL based approach to generate optimized CPU code. Finally, we also present experimental results for CPUs.}

\author{Karan Aggarwal}
\affiliation{%
    \department{Dept of CSA}
    \institution{Indian Institute of Science}
    \city{Bengaluru}
    \postcode{560012}
    \country{India}
}
\email{karan@iisc.ac.in}

\author{Uday Bondhugula}
\affiliation{%
    \department{Dept of CSA}
    \institution{Indian Institute of Science}
    \city{Bengaluru}
    \postcode{560012}
    \country{India}
}
\email{udayb@iisc.ac.in}


\begin{abstract}
Sparse matrix-vector multiplication (\textit{SpMV}) operations are commonly
used in various scientific and engineering applications. The performance of the
SpMV operation often depends on exploiting regularity patterns in the matrix.
Various representations and optimization techniques have been proposed to
minimize the memory bandwidth bottleneck arising from the irregular memory
access pattern involved. Among recent representation techniques, tensor
decomposition is a popular one used for very large but sparse matrices. Post
sparse-tensor decomposition, the new representation involves indirect accesses,
making it challenging to optimize for multi-cores and even more demanding for
the massively parallel architectures, such as on GPUs.

Computational neuroscience algorithms often involve sparse datasets while still
performing long-running computations on them. The Linear Fascicle Evaluation
(LiFE) application is a popular neuroscience algorithm used for pruning brain
connectivity graphs. The datasets employed herein involve the Sparse Tucker
Decomposition (STD) --- a widely used tensor decomposition method. Using this
decomposition leads to multiple indirect array references, making it very
difficult to optimize on both multi-core and many-core systems. Recent
implementations of the LiFE algorithm show that its SpMV operations are the key
bottleneck for performance and scaling. In this work, we first propose
target-independent optimizations to optimize these SpMV operations, followed by
target-dependent optimizations for CPU and GPU systems. The target-independent
techniques include: (1) standard compiler optimizations to prevent unnecessary
and redundant computations, (2)~data restructuring techniques to minimize the
effects of indirect accesses, and (3)~methods to partition computations among
threads to obtain coarse-grained parallelism with low synchronization overhead.
Then we present the target-dependent optimizations for CPUs such as:
(1)~efficient synchronization-free thread mapping, and (2)~utilizing BLAS calls
to exploit hardware-specific speed. Following that, we present various
GPU-specific optimizations to optimally map threads at the granularity of
warps, thread blocks and grid. Furthermore, to automate the CPU-based
optimizations developed for this algorithm, we also extend the PolyMage
domain-specific language, embedded in Python. Our highly optimized and
parallelized CPU implementation obtain a speedup of $6.3\times$ over the naive 
parallel CPU implementation running on 16-core Intel Xeon Silver (Skylake-based) 
system.  In addition to that our optimized GPU implementation achieves a speedup 
of $5.2\times$ over a reference optimized GPU code version on NVIDIA's GeForce 
RTX 2080 Ti GPU, and a speedup of $9.7\times$ over our highly optimized and 
parallelized CPU implementation.

\end{abstract}

\maketitle
\section{Introduction} \label{sec:introduction} Sparse matrix-vector
multiplication (\textit{SpMV}) is a key operation in many scientific and
engineering applications. As SpMV is typically memory bandwidth and latency
bound, it plays a significant role in determining the overall execution time as
well as the scalability of an application. Utilizing the architecture-specific
memory model to reduce its memory bandwidth requirement is a major challenge,
especially for highly parallel architectures such as GPUs, where exploiting the
regularity in unstructured accesses is key. Numerous prior works have been
proposed to improve the performance of SpMV, including that of the development
of new sparse representations~\cite{Mahmoud2018,bell2009sc,Sun2011},
representation-specific optimizations~\cite{bell2009sc,belgin2009ics,Guo2016}
and architecture-specific techniques~\cite{Baskaran2009OptimizingSM,
bell2009sc,liu2013ics,mellorcrummey2004,vuduc2005a,williams2007sc,wu2013ppopp,
shantharam2011}.

Tensor decomposition~\cite{kolda2009} is a popular technique to represent the
LHS matrix in SpMV as a combination of a tensor and other auxiliary data
structures in a way that drastically reduces the amount of storage. Tensor
decomposition has found use to perform SpMV operations efficiently across many
domains such as digital signal
processing~\cite{cichocki2015,sidiropoulos2017,DeLathauwer2007,DeLathauwer2008,
DeLathauwer2004}, machine learning~\cite{sidiropoulos2017}, data
mining~\cite{papalexakis2016,Sun2006,Sun2006sigkdd,Acar2005,Acar2006},
computational
biology~\cite{li2013,caiafa17nips,Mrup2006,Mrup2007,Mrup2008,Acar2007,
Acar2007embc, DeVos2007,MartnezMontes2004,Miwakeichi2004,Beckmann2005} and
several more mentioned by Kolda and Bader~\cite{kolda2009}.
Tucker~\textit{et~al.}~\cite{tucker1966} presented a widely used tensor
decomposition technique based on high-order singular value decomposition.
Tucker's technique is used in a range of
applications~\cite{zubair2013,yokota2014,perroscvs16,kolda2009}. More
importantly, the Tucker model is used to perform low-rank decomposition of
tensors to depict the sparse representations of matrices, and this is commonly
referred to as the Sparse Tucker Decomposition (STD)~\cite{tucker1966}. The
major challenge for an STD-based application however is that the sparse
representation entails multiple indirect array accesses. Therefore, efficiently
utilizing multi-core and many-core architectures poses a significant difficulty
because such accesses are both memory latency and bandwidth unfriendly.
However, employing STD for an SpMV operation is a necessary trade-off
considering the reduction in memory utilization obtained for a sparse matrix.

Building brain connectivity graphs or the wiring diagram of neural circuitry of
the brain, termed as \textit{connectome}, is an exciting computational
neuroscience conundrum involving large but sparse matrices. Understanding the
neural pathways is key to studying the connection between brain-regions and
behavior. Principally, a connectome can be described at various scales based on
the spatial resolution~\cite{Sporns2005,Wallace2004,Merboldt1985}. The scales
can be primarily categorized as microscale, mesoscale and
macroscale~\cite{keneddy2016}. A microscale connectome is a neuron-to-neuron
brain graph involving $10^{11}$ nodes (neurons) and $10^{17}$ edges (neuronal
connection); currently, obtaining and processing such large data appears
infeasible. A mesoscale connectome building technique is based on anatomical
properties of the brain, which again is not a viable choice due to poor
resolution of electron-microscopy~\cite{Kasthuri2009,Briggman2012}. Once
technology is enhanced, optimizing such large sparse datasets will still be a
formidable problem. In contrast, a macroscale level
connectome~\cite{Craddock2013} divides a brain model into 3D volumes called
\textit{voxels} (in the order of $10^6$ in number); this is thus a much more
tractable approach.

Diffusion-weighted Magnetic Resonance Imaging (\textit{dMRI}) is a popular
macroscale choice, that captures the diffusion of water molecules in the brain.
The dMRI along with \textit{tractography} techniques can be used to estimate
white matter connectivity in the human brain. These pathways represent physical
connections between brain regions and when analyzed in conjunction with
behaviour, can provide interesting insights into brain-behaviour relationships.
These insights are often essential in diagnosing diseases of the brain such as
Alzheimer's Disease~\cite{mueller2005adni}, a neurodegenerative disorder
involving degradation of white matter. While the non-invasive nature of dMRI
enables studying structural connectivity \textit{in-vivo} in humans, it suffers
from a major limitation in that the validity of the results cannot be tested
easily due to the lack of access to ground
truth~\cite{Jones2010,MaierHein2017}. Data acquisition protocols and
tractography approaches often depend on the specific scientific questions being
addressed and can differ significantly across cohorts. Thus, a standardized
evaluation technique to assess connectomes and establish evidence for
white matter pathways is critical for accurate and reliable estimation of
structural connectivity in the brain.

One such technique that addresses these shortcomings is the Linear Fascicle
Evaluation (LiFE)~\cite{pestilli2014,ccaiafa2017,caiafa17nips}, an algorithm
that prunes white matter connectomes to produce an optimized subset of fibers
that best explain the underlying diffusion signal. LiFE posits that the
diffusion signal in a voxel (a volume of brain tissue) can be approximated by a
weighted sum of the individual contribution of every streamline traversing that
voxel. The model thus entails a simple constrained optimization problem where
the weights associated with every streamline are estimated by minimizing the
error between the measured and predicted diffusion signal. This optimization is
carried out using a variant of the gradient descent method - the Subspace
Barzilai-Borwein non-negative least squares (SBBNNLS)
algorithm~\cite{dong13oms}, and involves iterative matrix multiplications.
However, large execution times and memory requirements have precluded the
large-scale use of the LiFE algorithm. While the memory issues have recently
been addressed with the use of sparse representations (Sparse Tucker
Decomposition~\cite{tucker1966}) of the data, the matrix-vector
multiplications, transformed to a more complex sequence of operations as
presented by Pestilli and Caiafa~\cite{ccaiafa2017} are still computationally
demanding, involving multiple indirect array accesses. Optimizing the
transformed SpMV operations on both multi-cores and GPUs is a challenging task
that is memory latency and bandwidth bound even for low-resolution dMRI
datasets.

In literature, several prior works have been proposed to tackle irregular
applications for both multi-core and GPU systems
such as~\cite{arenaz05pdpa, juan07ispdc,strout16parallelcomputing,
venkat14cgo,venkat15pldi,venkat2016sc}. These approaches use
\textit{inspector/executor} paradigm~\cite{arenaz05pdpa} to exploit regularity
in unstructured accesses. One such approach is presented by Venkat~\textit{et
al.}~\cite{venkat14cgo} to automate the code generation for a particular class
of application performing SpMV on GPUs. Other studies show various compiler
transformations to reduce the runtime overhead of code generation by the
inspector step in~\cite{venkat15pldi}, and generate optimized code for
wavefront parallelization for sparse-matrix representation
in~\cite{venkat2016sc}. These works have presented a semi-automatic approach to
analyze the data (using the inspector step) and then generate the optimized code
(using the executor step). Note that these works are limited to read non-affine
accesses. However, our work targets optimization of the SpMV operations of
LiFE, where the sparse matrix is decomposed using the STD technique. The new
representation of the matrix involves multiple irregular accesses which
includes both read as well as write array access. Therefore, due to presence of
such type of accesses, the exiting works will have a high runtime overhead.
However, in this work, we present a specific data restructuring method tuned
for LiFE with low run-time overhead. Furthermore, the prior works amortizes the
overhead due to inspector/executor across the iterations of a loop in a
program. In contrast, our work amortizes the overhead due to restructuring
across the several runs of the same program along with the iterations of a
loop. Additionally, our data restructuring optimization could potentially be
generalized and extended to other applications employing STD, although one
would have to look for similar or other data patterns. Thus, our work proposes
a tailored data restructuring method to tackles indirect access of SpMV
operations used in LiFE.

Prior works on optimizing the LiFE application considered distributed systems
and GPUs. Gugnani~\textit{et al.}~\cite{gugnani17hipc} proposed a distributed
memory based approach using MPI and OpenMP paradigms to parallelize the SpMV
operations of LiFE and obtained a speedup of $8.7\times$ over the original
approach. On the other hand, Madhav~\cite{madhav2017} developed a fast GPU
implementation to optimize the SpMV operations of LiFE by incorporating simple
optimization techniques. In another work, Kumar \textit{et
al.}~\cite{kumar2019} proposed a GPU-accelerated implementation for
\textit{ReAl-LiFE}~\cite{kumar2019}, a modification of LiFE application that
introduced regularized pruning constraint to build connectomes.

In this work, we optimize the SpMV operations by performing a number of
target-independent and target-dependent optimizations. The target optimizations
comprises: (1)~standard compiler optimizations, (2)~various data restructuring
methods, and (3)~techniques to partition computations among threads. These
optimizations can be automated and extended to other applications performing
SpMV operations where the matrix is decomposed using STD. The target-dependent
optimizations that we propose for multi-core architectures are following:
(1)~efficient synchronization-free thread mapping, and (2)~utilizing BLAS
calls, and for the GPUs the optimizations includes optimal techniques to map
threads at the granularity of warps, thread blocks and grids. Tailoring these
optimizations for the LiFE application, we obtain a speedup of $27.12\times$
for our highly optimized and parallelized CPU code over the original sequential
implementation, and speedups of $5.2\times$ and $1.87\times$ for our optimized
GPU implementation over a reference optimized GPU implementation (developed by
Madhav~\cite{madhav2017}) and over the ReAl-LiFE GPU implementation (tweaked to
perform same computations as the LiFE application) respectively. In addition,
our work can express the SpMV operation of LiFE in a high-level language and
abstract out other information using a domain-specific language (DSL) approach.
Using the domain information, we can perform optimizations that provide
significant improvements in performance and productivity. As a
proof-of-concept, we extend PolyMage~\cite{mullapudi15asplos}, a DSL designed
for image processing pipelines, to express the key matrix operations in LiFE
and automatically generate optimized CPU code to obtain similar performance
improvements compared to that of our hand-optimized CPU implementation.

The key contributions of this paper are as follows: 
\begin{itemize} 
	
\item We address challenges involved in optimizing SpMV operations of the LiFE
	application on multi-cores and GPUs by proposing various
		architecture-agnostic and architecture-dependent optimizations.

\item The target independent optimizations includes: (1)~standard compiler
	optimizations to avoid unnecessary and redundant computations, (2)~data
		restructuring methods to deal with multiple indirect array
		references that in turn make further optimizations valid and
		fruitful, and (3)~effective partitioning of computations among
		threads to exploit coarse-grained parallelism while avoiding
		the usage of an atomic operation.

\item The CPU-specific optimizations comprises: (1)~efficient
	synchronization-free thread mapping method to reduce load imbalance,
		and (2) mapping to BLAS calls to exploit fine-grained
		parallelism.

\item The GPU-specific optimizations include: (1)~leveraging fine-grained
	parallelism by utilizing a GPU's resources such as shared memory and
		the shuffle instruction, and (2)~effectively transforming loops
		to map iterations in a better way.

\item Then we present new constructs added to the PolyMage DSL to represent a
	sparse matrix and automatically generate optimized CPU code for the
		SpMV operations of the LiFE application.

\item We present experimental results and analysis to show the usefulness of
	the optimizations we incorporated for SpMV of LiFE, and also compare
		them with the existing implementations.

\item We present experimental results and analysis by varying various LiFE
	application parameters such as the number of voxels, number of fibers
		and different tractography techniques used to process the dMRI
		data for generating a connectome in the LiFE. 

\end{itemize}

The rest of this paper is organized as follows. We provide background on the
LiFE application in Section~\ref{sec:background}. We describe the problem and
challenges pertaining to optimizing SpMV computations of LiFE in
Section~\ref{sec:problemandchallenges}. The target-dependent and the
target-independent optimizations are described in
Section~\ref{sec:optimizations}. Then we present the constructs developed in
the PolyMage DSL to generate an optimized parallelized CPU code for the SpMV
operations in Section~\ref{sec:dsl}. Section~\ref{sec:experimentalevaluation}
presents details and analysis of experiments we performed by varying various
parameters of LiFE, the benefits of each optimization in an incremental manner,
and a comparison of various implementations of the SpMV. Related work is
discussed in Section~\ref{sec:relatedwork}, followed by conclusions and future
works in Section~\ref{sec:conclusions}. 
 
\section{Background} \label{sec:background} In this section, we introduce the
LiFE model, the optimization algorithm, the essential computations involved in
this algorithm as well as highlight the bottlenecks which have been addressed
in subsequent sections.  

\subsection{LiFE Algorithm} 
\begin{figure*}[!h]
\centering
\includegraphics[width=0.7\linewidth]{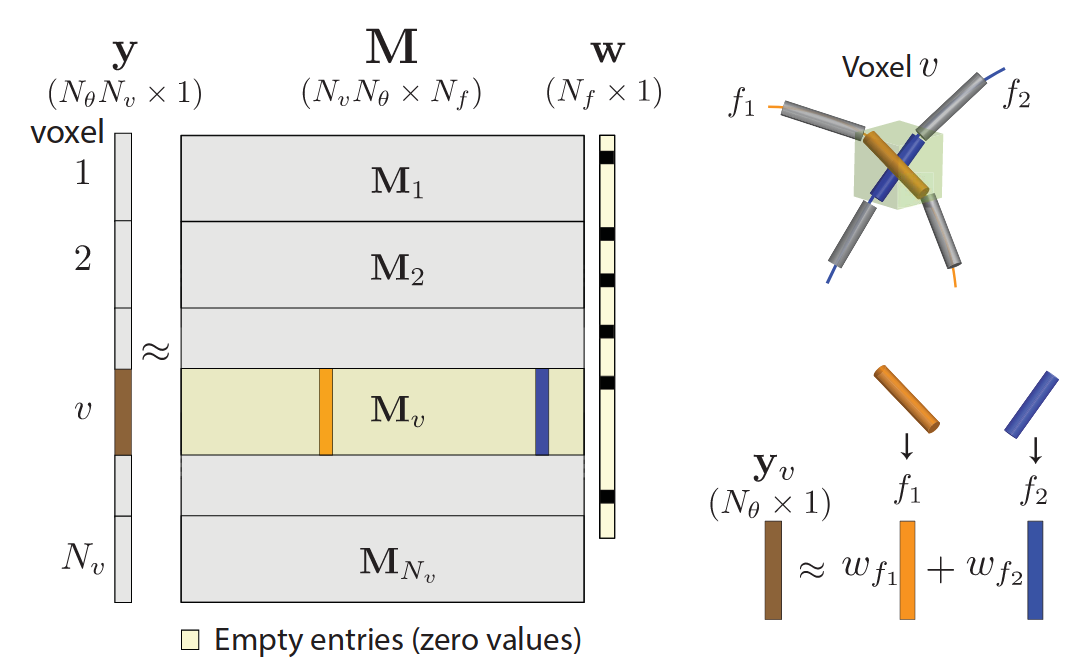}
\caption{SpMV operation in the LiFE algorithm [Source: Copyright 2017 by
	Caiafa~\textit{et~al.}~ 2017~\cite{ccaiafa2017} used under the CC BY
	4.0 license~\cite{license}.]} \label{fig:mwimage} 
\end{figure*}

Given a whole brain connectome obtained from
diffusion data, the goal of the LiFE is to retain only those fibers that best
predict the underlying diffusion signal. Let the total number of voxels in
which the signal is measured be $\bf N_{v}$. In each voxel, the signal is
obtained along multiple non-collinear gradient directions ($\bf N_{\theta}$),
and is represented by a vector $\bf y~\in~\mathbb{R}^{N_{\theta}N_{v}}$.
Further, the contribution of each fiber \textit{f} traversing voxel \textit{v}
is encoded in an array $\bf M~\in~\mathbb{R}^{N_{\theta}N_{v}\times N_{f}}$,
where $\bf N_{f}$ is the total number of fibers in the connectome. In each
voxel, \textit{v}, LiFE models the diffusion signal measured along each
gradient direction $\theta$ as the weighted sum of the contributions of every
fiber traversing \textit{v}. In other words, a candidate connectome is pruned
to obtain optimized connectome that best estimate the underlying diffusion
signal. Thus, the signal across all voxels and all gradient directions can be
summarized as: 
\begin{equation*} \scalebox{1.00}{$\bf y \approx Mw$} , \tag{1}
	\label{eq:1}
\end{equation*} 
\noindent where $\bf y\in\mathbb{R}^{N_{\theta}N_{v}}$ is a vector containing
demeaned diffusion signal for all voxels ($\bf v$) across all the gradient
directions ($\bf \theta$). Matrix $\bf M\in\mathbb{R}^{N_{\theta}N_ {v}\times
N_{f}}$, contains diffusion signal contribution by each fascicle ($\bf f$) at a
voxel ($\bf v$) in all diffusion directions ($\bf \theta$), and the $\bf w\in
\mathbb{R}^{N_{f}}$ vector contains the weight coefficients for each streamline
fascicle (Figure~\ref{fig:mwimage}). Equation~\ref{eq:1} is used to estimate
the weights by minimizing the error, is solved using following non-negative
least-squared optimization problem: 
\begin{equation*} \scalebox{1.00}{$\underset{\bf w}{\bf min}
	(\frac{1}{2}\left\Vert\left( \bf y-Mw \rm\rm\right) \right\Vert^2)$
such that $w_f \geqslant 0 , \forall f$}. \tag{2} \label{eq:2} 
\end{equation*}

\noindent The major challenge in solving Equation~\ref{eq:2} is the
significantly high memory requirements of the matrix $\bf M$. Even for small
datasets, $\bf M$ can consume about $40$GB. In another work, the authors of
LiFE proposed the \textit{ENCODE framework}~\cite{ccaifacode16}, wherein Sparse
Tucker Decomposition (STD)~\cite{tucker1966}, a sparse multiway decomposition
method to encode brain connectome, was used to reduce the memory consumption by
approximately $40\times$. Using the STD technique, the diffusion signal
contribution for a voxel ($\bf v$), $\bf M_{v} \in \mathbb{R}^
{N_{\theta}\times N_{f}}$ is represented as: \begin{equation*}
	\scalebox{1.00}{$\bf M_{v} = S_{0}(v) {D \Phi_{v}}$} , \tag{3}
\label{eq:3} \end{equation*} \noindent where $S_{0}(v)$ is the diffusion signal
measured in absence of gradient, $\bf D \in
\mathbb{R}^{N_{\theta}\times{N_{a}}}$ is a dictionary matrix for canonical
diffusion \textit{atoms} estimating individual streamline fiber based on their
orientation and signal contribution, and $\bf \Phi_{v} \in \mathbb{R}^{N_
{a}\times N_{f}}$ is a sparse binary matrix, whose column indicate primary
contributing atoms in individual fibers, in that voxel. Thus, an equation for
all $v$ can be re-written~as: 
\begin{equation*}
\scalebox{1.00}{$\bf Y = \Phi \times_{1} D \times_{2} S_{0} \times_{3} w^{T}$}
, \tag{4} \label{eq:4} 
\end{equation*} 
\noindent where $\Phi \times_{1} D \times_{2} S_{0}$ is 3D representation of
matrix $\bf M$ and $\bf \Phi$ is a 3D representation $\forall\bf~\Phi_{v}$,
with the goal to minimize the error between $\bf Y$ and $\bf y$ of
Equation~\ref{eq:1}. 

The optimization problem of Equation~\ref{eq:4} is solved using sub-space
Barzilie-Borwein non-negative least squares (SBBNNLS)
algorithm~\cite{dong13oms}. Typically, the SBBNNLS algorithm takes more than
$500$ iterations to converge, accounting for more than $92\%$ ($3$-$12$h) of
the total execution time of LiFE (for the original naive sequential C language
code).  Given $\bf w_0$ as the initial weight vector, for every iteration, the
weight vector is updated based on following equation: 
\begin{equation*} \scalebox{1.00}{$
w^{(i+1)} = [ w^{(i)} - \alpha^{(i)} \nabla g(w^{(i)})]_+$} , \tag{5}
\label{eq:5} 
\end{equation*}
\noindent where gradient, 
\begin{equation*} \scalebox{1.00}{ $\nabla g(w) =
M^T(Mw-y)$} , \tag{6} \label{eq:6} 
\end{equation*}
\noindent and the $\alpha^{(i)}$ step value for every even iteration is
computed using, 
\begin{equation*} \scalebox{1.00}{ $\alpha^{(i)} =  \dfrac
	{\langle \nabla \tilde{g}(w^{(i-1)}), \nabla \tilde{g}(w^{(i-1)})
	\rangle} {\langle M \nabla \tilde{g}(w^{(i-1)}), M\nabla
\tilde{g}(w^{(i-1)})\rangle} $}, \tag{7} \label{eq:7} 
\end{equation*}
\noindent and for the odd iterations using, 
\begin{equation*} \scalebox{1.00}{
		$\alpha^{(i)} =  \dfrac {\langle M \nabla \tilde{g}(w^{(i-1)}),
		M\nabla \tilde{g}(w^{(i-1)})\rangle} {\langle M^T M \nabla
		\tilde{g}(w^{(i-1)}), M^T M\nabla \tilde{g}(w^{ (i-1)})\rangle}
$}.  \tag{8} \label{eq:8} 
\end{equation*}
\noindent The Equations~\ref{eq:5}-\ref{eq:8} represent typical computations
necessary for SBBNNLS of LiFE, also shown in Algorithm~\ref{alg:sbb-nnls}.
Note that the tilde sign over gradient $\tilde{g}$ and "$+$" subscript in
Equation~\ref{eq:5} indicates projection to positive space, i.e., negative
values are replaced by zeros.
\begin{figure}[!t]
\centering
\begin{minipage}{.53\linewidth}
\begin{algorithm}[H]
\caption[SBBNNLS algorithm used in the LiFE algorithm]{SBBNNLS algorithm used in the LiFE algorithm \\(rewritten to 
represent matrix computations)}
\label{alg:sbb-nnls}
\begin{algorithmic}[1]
\State Given $\bf M$ as a connectome matrix, $\bf b$ as demeaned diffusion signal, and $\bf w^0$ (a vector) as initial approximation
\State \textbf{For} {$ i\gets 0, N-1$} 
\State The gradient descent method is performed to update weight vector using following computation:
\[\bf w^{(i+1)} = [ w^{(i)} - \alpha^{(i)} w^{'}]_+ \]
\State Gradient is calculated using: \[ \bf y = (M w^{(i)} - b )\] \[\bf  w^{'} = M^T y\]
\State The $\bf \alpha^{(i)}$ value is computed for different iterations as follows:
(a) \textbf{ODD} iteration: 
 \[\bf v^{'} = M w^{'}\]   
 \[\bf \alpha^{(i)} =  \dfrac{\langle w^{'}, w^{'}  \rangle} {\langle v^{'}, v^{'}\rangle}   \]
(b) \textbf{EVEN} iteration:
 \[\bf v^{'} = M w^{'}\]   \[\bf v^{''} = M^T v^{'}\]  
 \[\bf \alpha^{(i)} =  \dfrac{\langle v^{'}, v^{'} \rangle} {\langle v^{''}, v^{''} \rangle}   \]
\State \textbf{End For} 
\end{algorithmic}
\hrule
\footnotesize $\bf \langle v,v \rangle $~is a scalar dot product of a vector \textbf{$\bf v$}.\\ 
    $\bf '+'$ sign in subscript  indicates \textbf{$\bf w$} is projected to positive space. \\ 
    $\bf '\sim'$ sign over gradient indicates the gradient is projected to the positive space.
\end{algorithm}
\end{minipage}
\end{figure}

\subsection{Matrix Computations using Sparse Tensor Decomposition}
\label{subsec:matrix-computation} 
\begin{figure*}[!h]
	\centering 
	\subfloat[$ \bf y=Mw$
	performing diffusion signal computation (\texttt{DSC}).\label{fig:DSC}]{
		\includegraphics[height=4.75cm,scale=1]{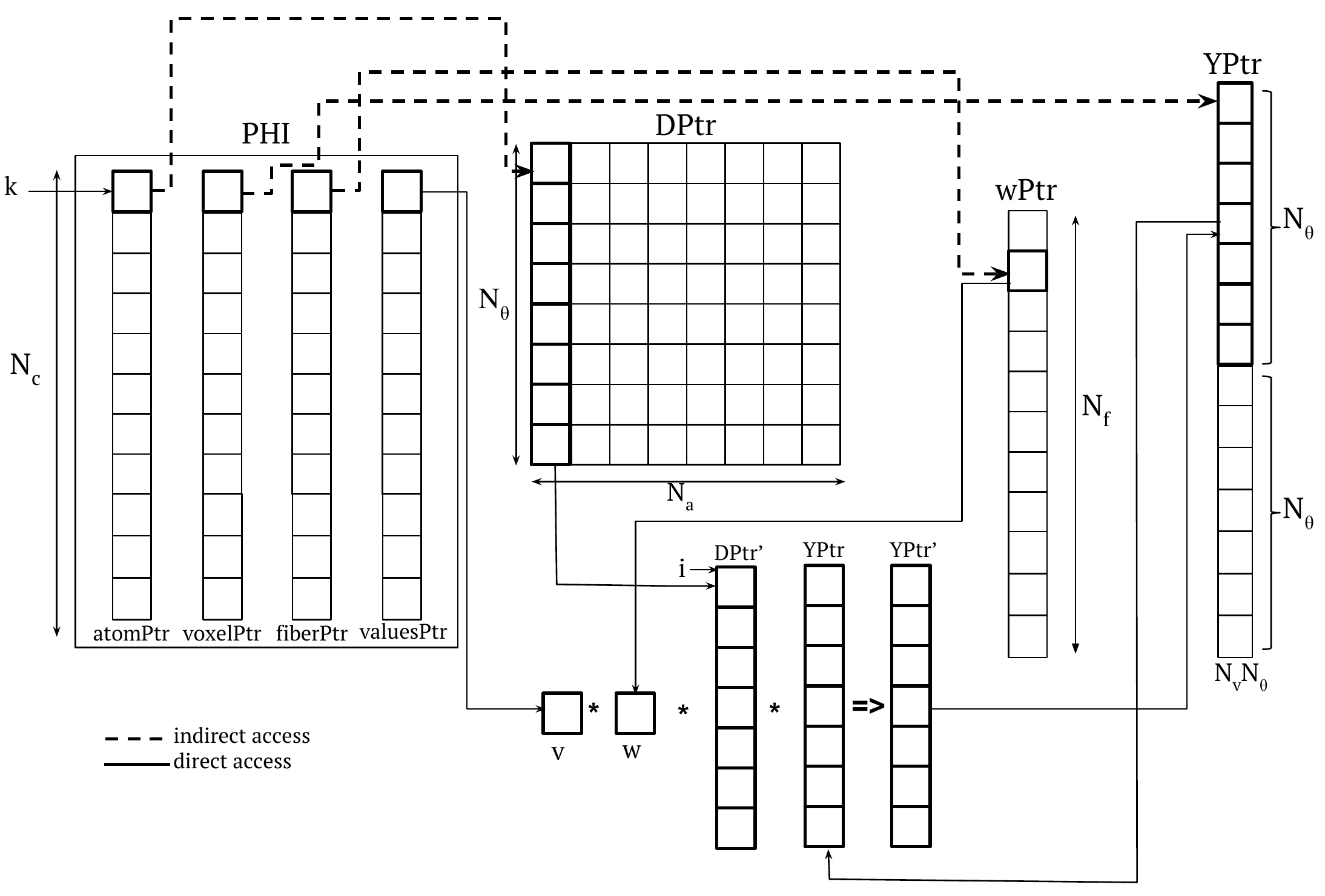} }
	\subfloat[$\bf w=M^Ty$ performing weight
	computation (\texttt{WC}).\label{fig:WC}]{
		\includegraphics[height=4.75cm,scale=1]{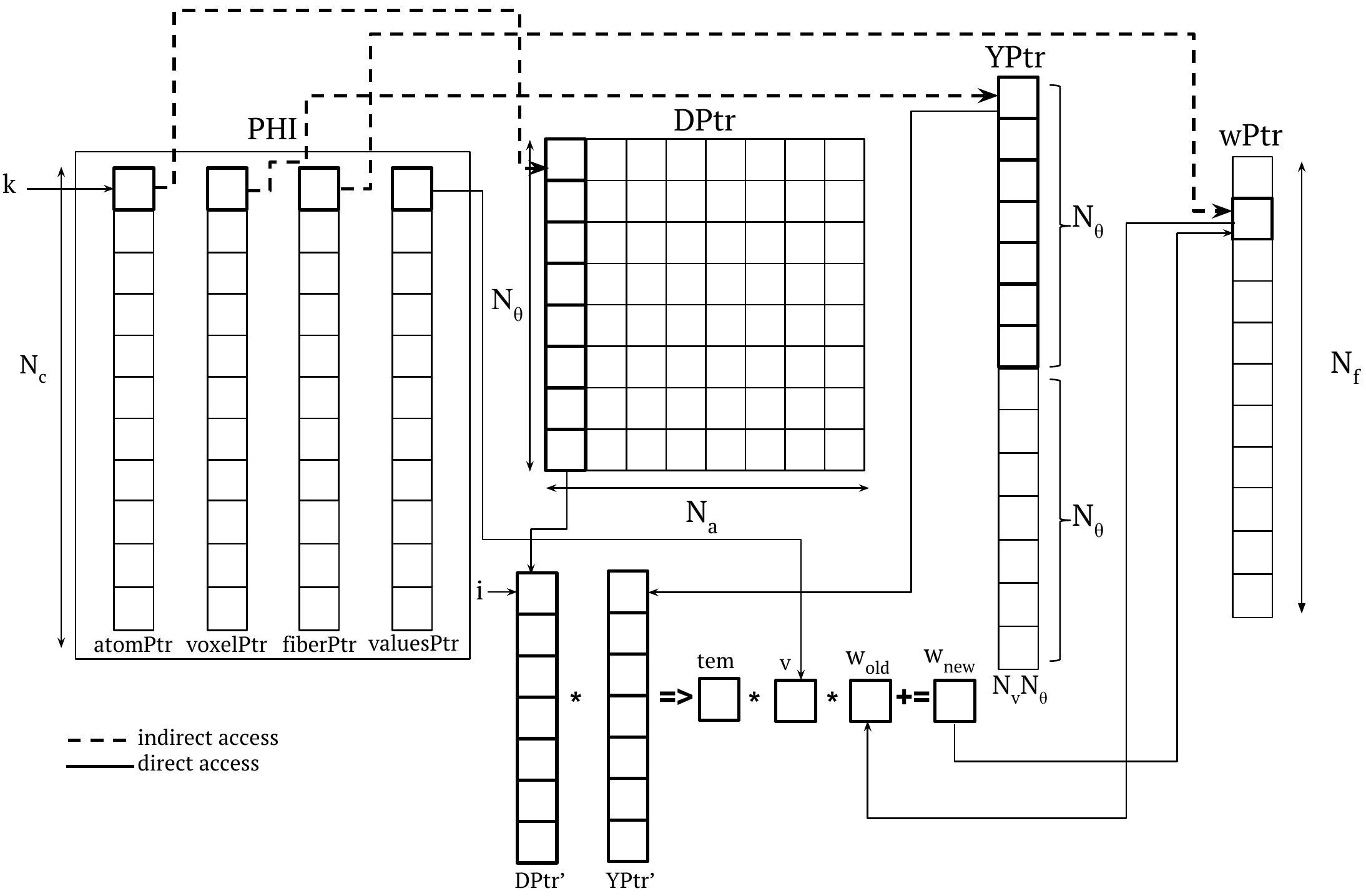} }
	\caption[Block diagrams of SpMV operations used in the LiFE
	algorithm]{Block diagrams of SpMV operations used in the LiFE
	algorithm} \label{fig:blockdiagram} 
\end{figure*}

The SBBNNLS algorithm involves two
compute-intensive SpMV operations involving the matrix $\bf M$, i.e., $\bf Mw$
and $\bf M^Ty$. On an average, every iteration (even or odd iteration) of
SBBNNLS requires the $\bf Mw$ operation twice and $\bf M^Ty$ $1.5$~times.  In
Figure~\ref{fig:blockdiagram}, it is shown how these simple SpMV operations are
transformed to a complex sequence of operations once the matrix $\bf M$ is
decomposed to a sparse format using STD.  The sparse tensor ($\bf \Phi$) stores
non-zero indices, (\textit{atomsPtr}, \textit{voxelsPtr} and
\textit{fibersPtr}), along with the values vector (\textit{valuesPtr}). In
Figure~\ref{fig:orig-code}, one can observe that the three indirection vectors
of the $\bf \Phi$ tensor --- \textit{atomsPtr}, \textit{voxelsPtr} and
\textit{fibersPtr}, redirects to the dictionary matrix \texttt{DPtr}, demeaned
diffusion signal vector \texttt{YPtr} and weight vector \texttt{wPtr}
respectively. The detailed algorithm for $\bf Mw$ and $\bf M^Ty$ matrix
operations are described in~\cite{ccaiafa2017}. The number of iterations of the
outermost loop depends on the number of coefficients ($\bf N_c$) representing
the non-zero indices in the $\bf \Phi$ tensor or the size of the
\textit{atomsPtr}/\textit{voxelsPtr}/\textit{fibersPtr} vectors. The number of
iterations of the innermost loop depends on the diffusion directions ($\bf
N_\theta$). Note that the innermost loop of $\bf Mw$ and $\bf M^Ty$ corresponds
to \texttt{daxpy} and \texttt{dot-product} operations respectively. It is also
important noting that the \texttt{wPtr} vector is projected to the positive
space; hence, the \texttt{wPtr} vector becomes sparser as it is updated after
the execution of each iteration of SBBNNLS (negative values are replaced by
zeros due to non-negativity property of SBBNNLS). 

\begin{figure*}[!h]
\centering 
\hfill \input{code/orig-mw} \hfill \input{code/orig-mtw}\hfill
\caption[Original sequential CPU code version of the SpMV operations]{Original
	sequential CPU code for the SpMV operations used in the LiFE algorithm
	\cite{ccaifacode16}.} 
\label{fig:orig-code} 
\end{figure*}
 
\section{Problem and Challenges} \label{sec:problemandchallenges} In this
section, we discuss problems and challenges associated with optimizing the SpMV
operations used in the SBBNNLS algorithm. 

\subsection{Large dataset} In Equations~\ref{eq:5}-\ref{eq:8}, we observe that
there are~two major SpMV operations involved, namely, $\bf y=Mw$ and $\bf
w=M^Ty$. The size of the matrix $\bf M$ depends on parameters such as the
number of voxels ($\bf N_v$), the number of fascicles ($\bf N_f$) and the
number of diffusion directions ($\bf N_\theta$). The number of diffusion
direction varies from $10$-$300$, voxels range from $10^5$ to $10^6$ and fibers
from $10^5$ to $10^7$; therefore, the memory consumption may range from a few
GBs to PBs. Thus, the matrix will typically not fit in commonly used memory
systems. The authors of the LiFE application analyzed the connectome matrices
and found that they are highly sparse in
nature~\cite{pestilli2014,ccaifacode16}. Hence, they proposed a low-rank Sparse
Tucker Decomposition (STD)~\cite{tucker1966} based approach to represent the
matrix $\bf M$ in a sparse tensor format and decompose it using domain-specific
information. After decomposition, a new challenge of multiple irregular
accesses is introduced, and this is discussed later in this section.

\subsection{Architecture-specific Challenges} We will discuss some
architecture-specific challenges posed in optimizing the SpMV operations of
SBBNNLS.

\paragraph*{\bf Multi-core architecture:} In multi-core architectures, the
processor can execute multiple independent instructions in parallel, hence
improving the speed of a program. Shared memory multi-core architectures uses a
multi-level cache memory to hide latency and reduce memory bandwidth
utilization.

\noindent \textit{Improving data reuse:} Shared memory multi-core architectures
uses multi-level cache memory to minimize the delay caused due to memory
latency. Hence, the data accessed multiple times should be reused optimally
before eviction from the cache memory.

\noindent \textit{Exploiting coarse-grained parallelism:} Coarse-grained
parallelism is splitting of large chunk of a program so that the communication
is minimized across the core. However, the coarse-grained parallelism requires
load balancing so that no core remains idle.

\noindent \textit{Exploiting fine-grained parallelism:} Fine-grained
parallelism is spitting small chunks of programs to facilitate load balancing.
However, faces a shortcoming of overhead caused due to usage of synchronization
barrier.

\paragraph*{\bf GPU architecture:} Modern GPUs are massively parallel,
multi-threaded, multi-core architectures with a memory hierarchy significantly
different from CPUs. Exploiting this parallelism and the various levels of the
memory hierarchy on a GPU is key to effectively optimizing the SpMV operations
of SBBNNLS.

\noindent \textit{Exploiting massive parallelism:} An appropriate partitioning
and mapping of threads to a thread block or a grid is essential to exploit the
massive parallelism on GPUs. One of the challenges here is to reduce the
overhead of communication across the thread blocks and warps/threads of a
thread block.

\noindent \textit{Efficiently using the GPU memory model:} The SMs of a GPU
share global memory, whereas local memory is allocated for a single thread.
Shared memory is used for sharing data among threads of a thread block. A GPU
provides multiple levels in its memory hierarchy to minimize the usage of
memory bandwidth.

\noindent \textit{Coalesced memory accesses:} Global memory accesses are
grouped such that consecutive threads access successive memory location. When
the threads of a warp access memory contiguously, the access is considered
fully coalesced otherwise considered partially coalesced access. Coalesced
memory accesses helps to reduce memory bandwidth requirement by loading local
memory in as few memory transactions.

\subsection{Indirect Array Accesses} \label{subsec:irregular}  As discussed in
Section~\ref{subsec:matrix-computation}, after STD-based tensor decomposition,
the SpMV operations of LiFE have several indirect array accesses. The
challenges that arises for CPUs due to unstructured accesses are following: (a)
the data reuse is low, hence memory bandwidth is poorly utilized, and (b) the
code is executed sequentially to avoid data races that occur due to the
dependent accesses. For GPUs, these irregular references (a) hinder the
utilization of massive parallelism of GPUs since synchronization and an atomic
operation is required to avoid data races, and (b) hamper the usage of various
fast GPU memory spaces and coalesced memory accesses. These are thus the main
challenges in optimizing the SpMV operations of the LiFE algorithm on
general-purpose multi-core and GPU systems.
 
\section{Optimizations} \label{sec:optimizations} 
\begin{figure}[!h] 
\centering
	\includegraphics[width=0.50\linewidth,scale=0.5] {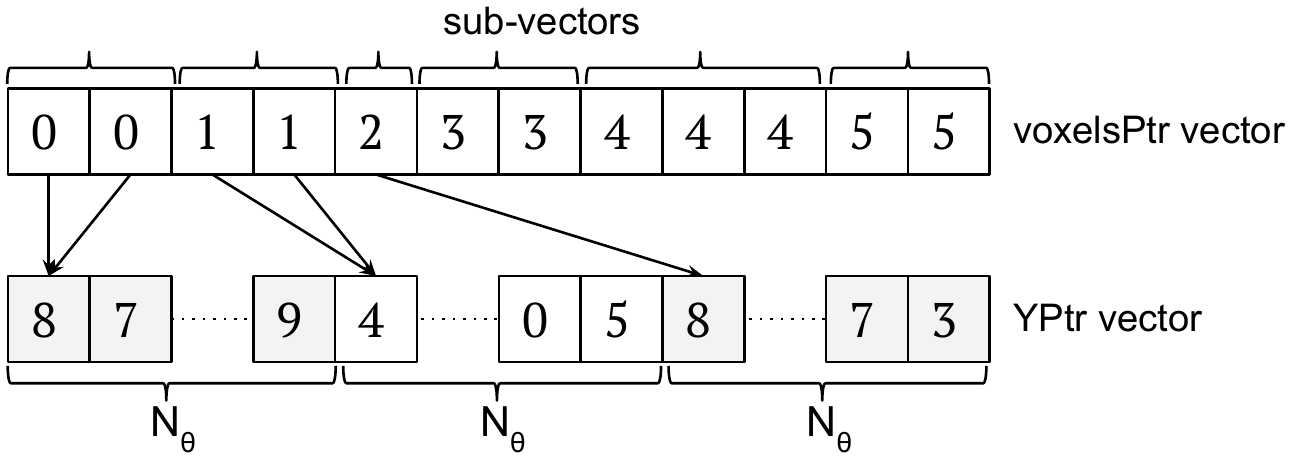}
	\caption[Sub-vectors of a vector]{Sub-vectors of the \textit{voxelsPtr}
indirection vector} \label{fig:subvec} 
\end{figure}

In this section, we discuss
details of the techniques we incorporate to optimize the SpMV operations used
in the LiFE algorithm. Firstly, we discuss target-independent optimization
techniques, followed by target-specific optimizations for parallel
architectures such as multi-core and GPU systems. We denote the SpMV operations
for computing the diffusion signal ($\bf y=Mw$) with $\bf \texttt{DSC}$ and the
weight ($\bf w=M^Ty$) with $\bf \texttt{WC}$. Also, in the discussion, wherever
we refer to a \textit{sub-vector of a vector} (Figure~\ref{fig:subvec}), it
corresponds to any contiguous part of a sorted indirection vector having the
same element value.

\subsection{Target-independent optimizations} \label{sec:tiopt} This section
introduces target-independent optimizations such as: (1)~basic compiler
optimizations to avoid unnecessary and redundant computations, (2)~various data
restructuring methods for inducing a potential regularity in the irregular
accessed data; also contributing to make further optimizations valid and
fruitful, and (3)~different ways to partition computations among parallel
threads to effectively exploit parallelism with low synchronization overhead.

\subsubsection{\bf Basic Compiler Optimizations:} In this sub-section, we discuss some
of the standard compiler optimizations that we incorporate to obtain trivial
performance improvement.

\paragraph*{Removing redundant computation:} The dictionary matrix
(\texttt{DPtr}) and the demeaned diffusion signal matrix (\texttt{YPtr}) are
used in the vector format for the SpMV operations (refer to Figure
\ref{fig:orig-code}). Therefore, to compute the actual offset of these vectors,
we multiply the number of diffusion direction ($\bf N_\theta$) with the
elements of the \textit{atomsPtr} and \textit{voxelsPtr} indirection vectors.
The original sequential CPU code computes the actual offset for every iteration
of the SpMV operations of the SBBNNLS algorithm.  However, we removed this
redundant computation, by computing it one time before the start of SBBNNLS in
the MATLAB code of LiFE. This reduced the overhead of computing the actual
offsets for \texttt{DPtr} and \texttt{YPtr} in the SpMV computations.

\paragraph*{Loop-invariant code motion:} Loop-invariant code motion
optimization is utilized when a code fragment performs the same operation and
computes the same output value for the different iterations of a loop, then
that code fragment is hoisted out of the loop. In LiFE, the \texttt{DSC}
operation computes the product of the weight vector (\texttt{wPtr}) and the
values vector (\texttt{valuesPtr}), which remains same for the innermost loop
of SpMV operations. Hence, this code fragment is hoisted out and its result is
stored in a temporary variable to utilize it across the several iterations of
the loop. Thus, this optimization reduced the overhead of computing the
invariant-code from several times for the innermost loop to one time.

\paragraph*{Strength reduction for arrays:} Some expressions that take
more memory and CPU cycles to execute, can be compensated by an equivalent
though less expensive expression. In LiFE application, the indirection vectors
such as \textit{atomsPtr}, \textit{voxelsPtr} and \textit{fibersPtr} are stored
and passed as a double precision data type, and used as an index (after
explicit type conversion to integer) for the \texttt{DPtr}, \texttt{YPtr} and
\texttt{wPtr} vectors respectively. Thus, to reduce memory consumption and
exploit a less expensive expression for these double precision indirection
vectors, they are casted to the integer data type. This optimization is
incorporated before the start of SBBNNLS in the MATLAB code and utilized across
the several iterations of SBBNNLS. In addition to that, this optimization
helped to cut down the data transfer overheads on GPUs due to the reduced size
of the indirection vectors. \\

These simple and straightforward optimizations can be incorporated for both the
\texttt{DSC} and \texttt{WC} operations without much effort.

\subsubsection{\bf Data Restructuring:} \label{subsec:ds} The LiFE algorithm is highly
irregular due to the presence of multiple indirectly accessed arrays. In
Figure~\ref{fig:blockdiagram}, we observe that due to the STD-based
representation of the matrix $\bf M$ in SpMV, three indirection vectors are
involved --- \textit{atomsPtr}, \textit{voxelsPtr} and \textit{fibersPtr},
redirecting to the \texttt{DPtr}, \texttt{YPtr} and \texttt{wPtr} vectors
respectively. These indirect array accesses procure low data reuse and prove to
be a major hindrance in code parallelization as well; thus, they are a major
bottleneck in optimizing the SpMV.

After analyzing the sparse datasets of LiFE, we observe that there exist
several element values of an indirection vector redirecting to the same
location of an indirectly accessed vector. Therefore, this is a potential
source to exploit data locality. To utilize this property of the sparse
datasets, we restructure the \texttt{Phi} tensor ($3$-D sparse representation
of $\bf M$, represented by $\bf \Phi$) data based on an indirection vector to
leverage regular data access patterns. If the $\bf \Phi$ tensor is restructured
based on one of the indirection vectors (for example \textit{voxelsPtr}), then
the other indirection vectors (such as \textit{atomsPtr} and
\textit{fibersPtr}) are accessed irregularly. Hence, a major challenge in
optimizing this irregular application is to identify a near-optimal method to
restructure with low runtime overhead. Thus, to achieve high performance for an
SpMV operation, we determine the data restructuring to be incorporated at
runtime based on the choice of a dimension (such as atom, voxel or fiber). We
now discuss different data restructuring choices coupled with their strengths
and weaknesses.

\paragraph*{Atom-based Data Restructuring:} In the atom-based data
restructuring method, we sort the \textit{atomsPtr} vector, and depending on
that, the $\bf \Phi$ tensor is restructured by reordering the voxel, fiber, and
values dimensions. This method captures data reuse for the dictionary vector
\texttt{DPtr} in both the \texttt{DSC} and \texttt{WC} operations; but it leads
to poor data reuse along the other two indirectly accessed dimensions, that is,
voxel and fiber.

\paragraph*{Voxel-based Data Restructuring:} In the voxel-based data
restructuring method, we sort the \textit{voxelsPtr} vector, and depending on
that, the $\bf \Phi$ tensor is restructured by reordering the atom, fiber, and
values dimensions. This data restructuring method captures data reuse for the
demeaned diffusion signal vector \texttt{YPtr} in the \texttt{DSC} and
\texttt{WC} operations; but it leads to poor data reuse along the other two
indirectly accessed dimensions, atom and fiber.

\paragraph*{Fiber-based Data Restructuring:} In the fiber-based data
restructuring method, we reorder \textit{fibersPtr}, and depending on that, the
$\bf \Phi$ tensor is restructured by reordering the atom, voxel, and values
dimensions. The fiber-based approach captures data reuse for the \texttt{wPtr}
vector. However, this approach loses a chance to capture data reuse for the
vectors \texttt{YPtr} and \texttt{DPtr}. By inspection we found that
\texttt{YPtr} and \texttt{DPtr} vectors captures a much better regular data
access pattern compared to \texttt{wPtr}.  Thus, we skip the fiber-based data
restructuring for further analysis.

\paragraph*{Hybrid Data Restructuring:} Hybrid data restructuring
technique is a merger of the atom-based and the voxel-based data restructuring
methods. In this technique, we first execute the \texttt{DSC} and \texttt{WC}
operations for both the atom-based and the voxel-based restructuring method
three times, and based on the average execution time, we select a dimension
that achieves better performance for an SpMV operation. Therefore, we obtain
data reuse along the atom dimension or the voxel dimension. Then, the $\bf
\Phi$ tensor is restructured again by reordering the sub-vectors of the
selected dimension, to capture a chance of data reuse along the other dimension
(that is, other than the selected dimension). This technique will be useful for
very large datasets. However, currently for this method, the performance
improvement is almost negligible due to the data access patterns of the
low-resolution datasets used by us and additionally, this technique has a high
overhead of an additional data restructuring. Hence, we skip the hybrid-based
restructuring for further evaluation as we use only low-resolution datasets
(having small memory utilization) for our evaluation.\\

Another advantage of data restructuring besides from that of significant
improvements in data reuse due to regular accesses is that the other
optimizations to exploit parallelism and reduce synchronization overheads
(discussed later in this section) become valid and profitable. Therefore, data
restructuring play a key role to optimize the SpMV operations of LiFE.

The data restructuring to be incorporated is dependent on the input dMRI data
and other parameters (such as the number of voxels and fibers) along with a
tractography algorithm used.  Therefore, we automate the determination of the
data restructuring at runtime, by choosing a technique having lower average
execution time for three runs. We included the data restructuring optimization
in the LiFE algorithm's MATLAB implementation before invoking the SBBNNLS
algorithm, so that the overhead ($3$-$5\%$ of the total execution time of
SBBNNLS) is amortized across several iterations of the non-negative
least-squared algorithm. Note that for a different architecture and an SpMV of
LIFE, the data restructuring technique that obtains a near-optimal performance
may vary.

\subsubsection{\bf Computation Partitioning:} \label{subsec:tb} Post data
restructuring, the other problem in improving performance of the SpMV
operations was the usage of an atomic operation, which was required due to
parallel threads performing a reduction in the \texttt{DSC} and \texttt{WC}
operations (Figure~\ref{fig:orig-code}). This causes a high synchronization
overhead at runtime, detrimental to the exploitation of massive parallelism on
multi-cores and GPUs.  We note that the communication among threads can be
reduced by mapping computations of the outermost loop of SpMV to a single
thread based on the coefficient ($\bf N_c$) parameter of the LiFE, or on the
\textit{atomsPtr} or the \textit{voxelsPtr} dimension. Thus, another major
challenge in optimizing the SpMV operations is to determine a method to
partition computations for effectively exploiting parallelism and further
improving the data reuse for the \texttt{YPtr} and \texttt{DPtr} vectors. We
discuss various approaches to handle the computations performed by each thread
block in addition to their merits and demerits in detail.

\paragraph*{Coefficient-based computation partitioning:} In the
coefficient-based computation partitioning technique, a single thread handles
computations of a single coefficient or in other words single non-zero value of
the sparse tensor ($\bf \Phi$).  The parallelism provided by multi-cores and
GPUs can be effectively used by the coefficient-based technique, but this leads
to a loss of data reuse for the \texttt{YPtr} and \texttt{DPtr} vectors.
Additionally, as stated in Section~ \ref{subsec:matrix-computation}, the
\texttt{wPtr} vector is projected to positive space, implying that the negative
values are replaced by zeros. This sparse property of \texttt{wPtr} is
particularly useful for the \texttt{DSC} operation as a lot of unnecessary
computations can be avoided. However, this computation partitioning technique
requires usage of an atomic operation due to the reduction of the \texttt{YPtr}
and \texttt{wPtr} vectors in the \texttt{DSC} and \texttt{WC} operations
respectively. The coefficient-based technique also hinders incorporation of
certain other optimizations discussed later in this section.

\paragraph*{Atom-based computation partitioning:} In the atom-based
computation partitioning technique, computations are partitioned across the
threads based on the atom dimension, where each thread handles computations of
a particular atom. Therefore, this technique obtains good data reuse for
\texttt{DPtr} but lose an opportunity to exploit data reuse for \texttt{YPtr}.
Note that the atom-based computation partitioning uses the atom-based data
restructuring. 

\paragraph*{Voxel-based computation partitioning:} In the voxel-based
partitioning technique, computations are partitioned across the voxels, where
each thread handles computations of one voxel. In this way, the voxel-based
partitioning obtains excellent data reuse for \texttt{YPtr} (as it is accessed
twice due to reduction) but lose an opportunity to exploit data reuse for
\texttt{DPtr}. Note that the voxel-based computation partitioning uses the
voxel-based data restructuring. \\

The disadvantage of using the atom-based and the voxel-based techniques are (1)
all iterations associated with a sub-vector of voxel or atom dimension are
executed sequentially; therefore, this leads to a loss to fully utilize the
sparse property of \texttt{wPtr}, and (2) each thread block handles several
iterations depending on the size of a sub-vector, where the size may vary from
one to thousands of iterations; hence, this induces load imbalance. Therefore,
due to the moderate parallelism of multi-core CPUs, the load imbalance might be
more prominent in them.  Thus, to tackle the load imbalance in CPUs, we propose
a new technique discussed later in Section~\ref{subsec:parallel}. However, on
GPUs, the load imbalance issue does not impact much because the number of
iterations of the outermost loop ($\bf N_c$) in the SpMV operations is
extremely large compared to the maximum possible thread blocks that can be
scheduled to even the modern GPUs.

Therefore, this optimization helped in exploiting coarse-grained parallelism
with excellent data reuse. We also observed that avoiding the atomic operation
improves the performance considerably than taking advantage of the sparse
property of \texttt{wPtr}. Thus, by performing experiments on the datasets used
by us, we found that for \texttt{DSC} the coefficient-based partitioning is
favourable for CPUs and the voxel-based partitioning is favourable for GPUs,
whereas for \texttt{WC} the coefficient-based technique is favourable for both
CPUs and GPUs.

\subsection{Target Specific Optimizations} In this sub-section, we present
target-specific optimization techniques to optimize SpMV operation of LiFE on
multi-core and GPU architectures.

\subsubsection{\textbf{CPU-specific Optimizations:}} \label{sec:cpuoptimizations}
Firstly,
we discuss benefits and applicability of incorporating target-independent
optimizations on CPUs. Then we introduce CPU-specific optimizations such as
efficient synchronization-free thread mapping to utilize coarse-grained
parallelism with reduced load imbalance and usage of BLAS library calls to
exploit fine-grained parallelism.

\paragraph{\bf Target-independent optimizations on CPUs:} \label{subsec:tiopt-cpu} In
Section~\ref{sec:tiopt}, we discussed three target-independent optimizations
for SpMV operations of LiFE. The basic compiler optimizations presented are
directly applicable to obtain trivial performance improvement on CPUs. The data
restructuring optimization helped to enhance data reuse for \texttt{YPtr} and
\texttt{DPtr} vectors in SpMV operations, and further assisted to validate
parallelism. Next, we presented different ways to partition computations among
the parallel threads to exploit coarse-grained parallelism. However, this
optimization aggravated the issue of load imbalance for atom-based and
voxel-based partitioning, and an issue of high synchronization overhead for the
coefficient-based partitioning due to the usage of an atomic operation to avoid
data races. It is difficult to improve the load balance for the atom-based and
voxel-based partitioning methods; however, for the coefficient-based
partitioning, the overhead issue can be addressed if the atomic operation is
evaded. Hence, to tackle this issue we propose a CPU-specific optimization,
which is discussed next in this section.

\paragraph{\bf Efficient Synchronization-free Thread Mapping:}
\label{subsec:parallel} Earlier in Section~\ref{subsec:tb}, we discussed
various ways to partition computations to the parallel threads. We concluded
that for both the SpMV operations, the atom-based and the voxel-based
partitioning techniques were not profitable due to the load imbalance issue. In
addition to that, the atom-based and voxel-based methods required an atomic
operation due to the reduction of \texttt{YPtr} vector in \texttt{DSC}
operation and \texttt{wPtr} vector in \texttt{WC} operation respectively.
Whereas, the coefficient-based did not have a prominent load imbalance issue
but still it was not profitable due to the usage of an atomic operation.

For \texttt{WC} operation, we observe that for different computation
partitioning techniques, the performance is influenced due to the usage of an
atomic operation for the reduction of \texttt{wPtr}; although, based on
experiments we discovered that the usage of the atomic operation did not
deteriorate the performance much. We found that coefficient-based partitioning
is the best choice among the other methods because it exhibits a much better
load balance.  However, for \texttt{DSC}, we observed that there was a
significant drop in performance due to the usage of atomic operation (for all
the partitioning methods) and the load imbalance issue (for atom and voxel
based methods). Thus, using the coefficient-based partitioning method, we
tackle this issue by proposing an efficient synchronization-free thread mapping
technique to exploit coarse-grained parallelism without the usage of an atomic
operation to improve the performance of \texttt{DSC}.

\begin{figure*} \centering \subfloat[atom-based data
	restructuring\label{fig:atombased}]{
		\includegraphics[width=0.45\linewidth,scale=1]{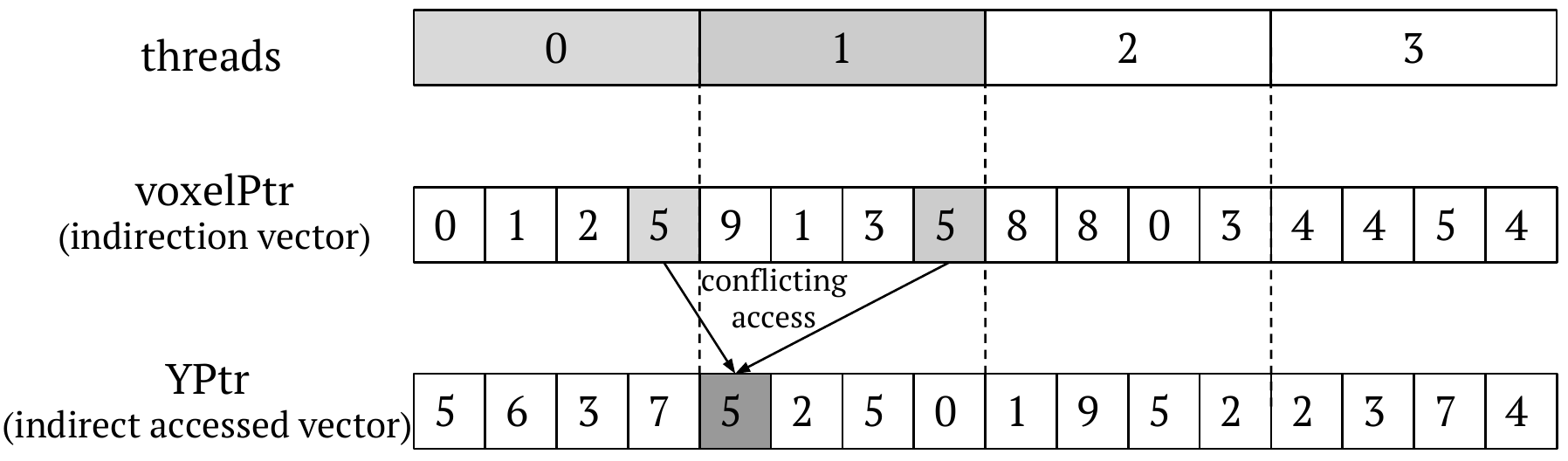}
		} \subfloat[voxel-based data
		restructuring\label{fig:voxelbased}]{
			\includegraphics[width=0.45\linewidth,scale=1]{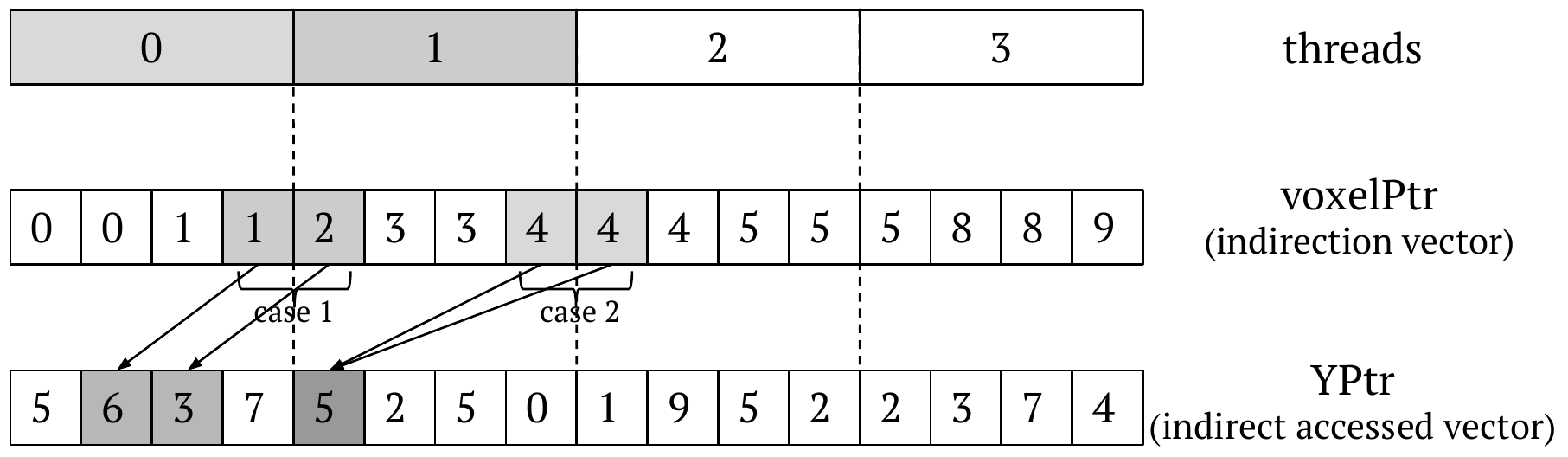}
			} \caption[Code parallelization validity for various
			data restructuring techniques]{The diagram represents
			the \textit{atomsPtr} unstructured vector redirects to
			the \texttt{YPtr} vector. (a)~\textit{voxelPtr}
			indirection vector is irregular when reordered based on
			the atom dimension (b)~\textit{voxelPtr} indirection
			vector is structured when reordered based on voxel
			dimension. The case 1 represents a sub-vector of the
			\texttt{voxelPtr} scheduled completely to a single
			threads.  Whereas, the case 2 represents a sub-vector
			of the \texttt{voxelPtr} split across the two threads.}
			\label{fig:parallel} \end{figure*}

In Figure~\ref{fig:parallel}, we observe the usage of coefficient-based
splitting technique for the different data restructuring methods for
\texttt{DSC}. Figure~\ref{fig:atombased} shows the atom-based restructuring
technique reorders the \textit{voxelsPtr} vector in such a way that there are
high chances of data race at runtime; hence, this method exhibits poor
performance due to requirement an atomic operation to avoid data race.
Figure~\ref{fig:voxelbased} shows that the voxel-based technique has a low
chance of data dependence but cannot be eliminated completely; hence, this
technique too requires an atomic operation. However, we found that there only
two instances might occur for a sub-vector of the \textit{voxelsPtr} vector
when the voxel-based data restructuring method is employed. These instances
are: (1) the entire sub-vector is scheduled to the same thread; hence, it
causes no issue due to sequential execution of the iterations of the sub-vector
(\texttt{case 1} of Figure~\ref{fig:voxelbased}), and (2) the sub-vector is
split across the two threads (\texttt{case 2} of Figure~\ref{fig:voxelbased});
therefore, for this case an atomic operation is required due to a chance of
data dependence at run-time.

To tackle this issue, we ensure that the sub-vector of the \textit{voxelsPtr}
vector is scheduled to the same thread with a low load imbalance. In
Figure~\ref{fig:voxelbased}, we can observe in the \texttt{case 2} that the
sub-vector (with $4$ value) is split across the threads $1$ and $2$. To avoid
any chance of occurrence of conflicting access, the sub-vector has to be
scheduled to either of the one thread. If the sub-vector is scheduled to the
\texttt{thread-1} then it will compute two additional computations, whereas if
the sub-vector is scheduled to the \texttt{thread-2} then it will compute only
one additional computation. Hence, scheduling the sub-vector to the
\texttt{thread-2} will help to reduce the load imbalance. The small overhead of
load imbalance is a necessary trade-off considering the reduction in execution
time obtained for parallel execution of the \textit{DSC} operation without the
usage of an atomic operation.

Thus, to exploit the coarse-grained parallelism for the \textit{DSC} operation
without atomic operation and with reduced load imbalance, we proposed an
efficient synchronization-free thread mapping using the coefficient-based
partitioning and the voxel-based data restructuring method.

\paragraph{\bf Mapping to BLAS calls:}  \label{subsec:blas} Basic linear
algebra subroutines (BLAS)\blfootnote{Usage of BLAS calls on Intel platforms
have a slightly different result on different runs of the same program due to
rounding error.  https://github.com/xianyi/OpenBLAS/issues/1627} packages
are often hand-optimized to obtain close to peak performance on various
hardware. It is thus useful to leverage these automatically in a DSL setting.
We make use of optimized BLAS call in the SpMV operations of the SBBNNLS
algorithm. BLAS call improved the overall performance of the LiFE algorithm
significantly. We discuss usage of a BLAS call in each of the SpMV operations
of SBBNNLS.

\noindent\textit{BLAS call for \texttt{DSC} operation:} The code fragment in the
innermost loop of \texttt{DSC} (refer to Figure \ref{fig:orig-mw}) corresponds
to scalar-vector product. We substitute the code fragment with the
\texttt{daxpy} BLAS call to obtain significant performance improvement. In the
BLAS call, dictionary vector (\texttt{DPtr}) is used as an input vector and the
product of a value in the weight vector (\texttt{wPtr}) and the values vector
(\texttt{valuesPtr}) is used as a scalar input. The output is used to update
the demeaned diffusion signal vector (\texttt{YPtr}).

According to the SBBNNLS stated in Algorithm~\ref{alg:sbb-nnls}, \texttt{wPtr}
is projected to the positive space; hence, due to this property of
\texttt{wPtr} the negative values are replaced by zeros. Therefore, the
\texttt{wPtr} vector is sparse in nature. Hence, in the \texttt{DSC} operation,
if the scalar value obtained from the product vector \texttt{wPtr} and vector
\texttt{valuesPtr} is zero then invoking the BLAS call is futile and should be
avoided to refrain from unnecessary computations.

\noindent\textit{BLAS call for the \texttt{WC} operation:} The code fragment in the
innermost loop of \texttt{WC} (refer to Figure \ref{fig:orig-mtw}) corresponds
to vector-vector dot product. We substitute the code fragment with the
\texttt{dot} BLAS call to obtain performance improvement. In \texttt{dot} BLAS
call, the \texttt{YPtr} and \texttt{DPtr} vectors are used to update the
\texttt{wPtr} vector. However, in contrast to the \texttt{DSC} operation, the
execution time remains almost the same throughout SBBNNLS.

\noindent Usage of BLAS call provided fine-grained parallelism for the SpMV
operations and improved the performance considerably. Particularly, the
\texttt{DSC} operation was greatly benefited by the usage of the BLAS call. \\

To summarize the optimization of SpMV on CPUs, first we performed the
target-independent optimizations, followed by the CPU-specific optimizations to
obtain a highly optimized CPU code for the SpMV operations of SBBNNLS. We also
extended the PolyMage DSL to incorporate all the optimization presented in this
section to automatically generate optimized parallelized code involving the
sparse representation of the SpMV operations of SBBNNLS and obtained comparable
performance to that of the manually optimized version (\textit{CPU-opt}). We
will discuss more on the DSL extension in Section~\ref{sec:dsl}. Note 
that some of the CPU optimizations require runtime data analysis such as the 
optimization presented in Section~\ref{subsec:parallel}. Thus, it could 
not be incorporated for the automated CPU code version and as a result the automated 
code version could not achieve the similar performance compared to that of the 
hand-optimized CPU code version.

\subsubsection{\textbf{GPU-specific Optimizations:}} \label{sec:gpuoptimizations} Firstly,
we discuss benefits and applicability of incorporating target-independent
optimizations on GPUs. Then, we present various GPU-specific optimizations to
optimally map threads at the granularity of warps, thread blocks and grid to
obtain fine-grained parallelism and improved data reuse. We use GPU code
developed by Madhav~\cite{madhav2017}, shown in Figure~\ref{fig:ref-code}, as a
reference GPU code version. 

\paragraph{\bf Target-independent optimizations on GPUs:} In
Section~\ref{sec:tiopt}, we discussed a number of target-independent
optimizations for SpMV operations. For GPUs, the basic compiler optimizations
presented is useful to obtain minor performance. The data restructuring
optimization proposed captured enhanced data reuse for \texttt{YPtr} and
\texttt{DPtr} vectors, and further aided to legitimize parallelism. Following
that, we presented different ways to partition computations among the parallel
threads to exploit coarse-grained parallelism. However, this optimization had
similar issues for a GPU that we discussed in Section~\ref{subsec:tiopt-cpu} for a
CPU; although, the issue of the load balance discussed earlier for a CPU is not
prominent for a GPU due to its massive parallelism. Thus, we do not introduce
any new optimization to tackle load imbalance issue for the GPUs and take a
step forward to exploit fine-grained parallelism in the SpMV operations.

\begin{figure*}[t] \hfill \input{code/ref-mw} \hfill \input{code/ref-mtw}
\hfill \caption{Reference GPU code for the SpMV operations used in the SBBNNLS
algorithm} \label{fig:ref-code} \end{figure*}

\begin{figure*} \centering \subfloat[\label{fig:main1a}]{
		\includegraphics[width=0.42\linewidth,scale=0.5]{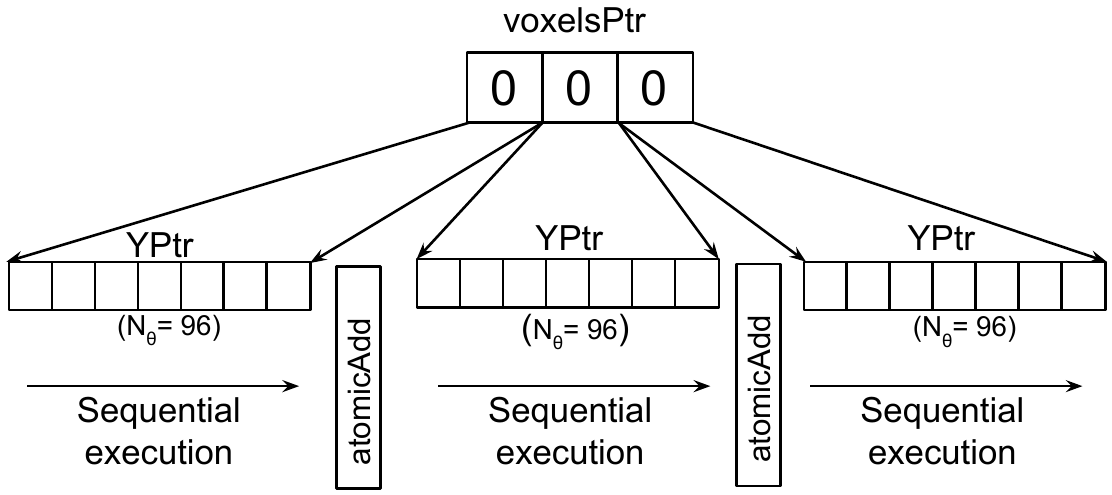}
		} \subfloat[\label{fig:main1b}]{
			\includegraphics[width=0.32\linewidth,scale=0.5]{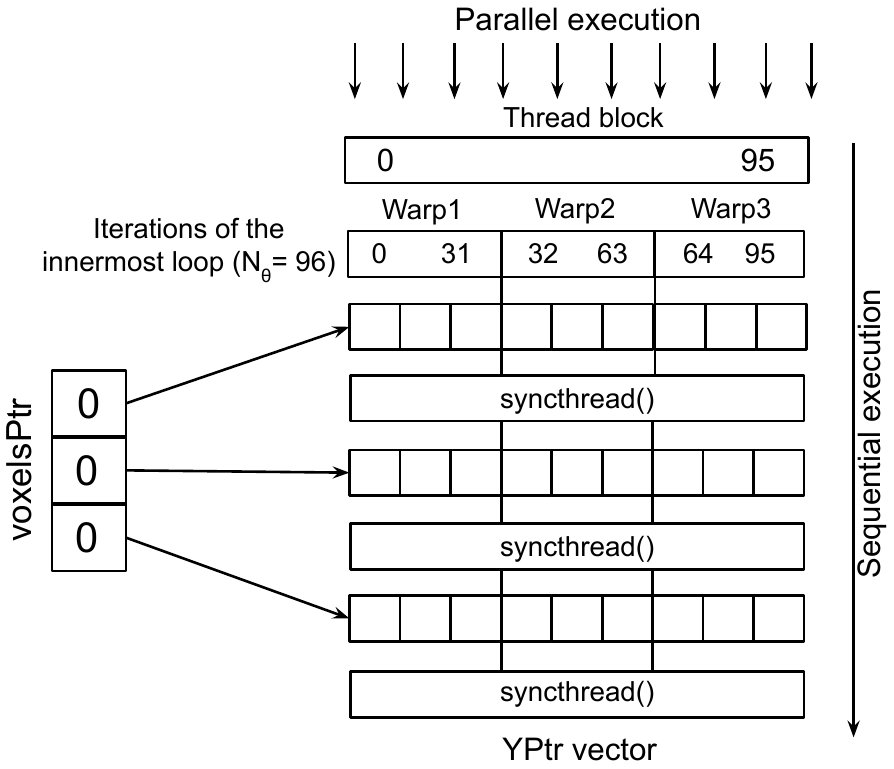}
			}

    \subfloat[\label{fig:main1c}]{
	    \includegraphics[width=0.3\linewidth,scale=0.5] {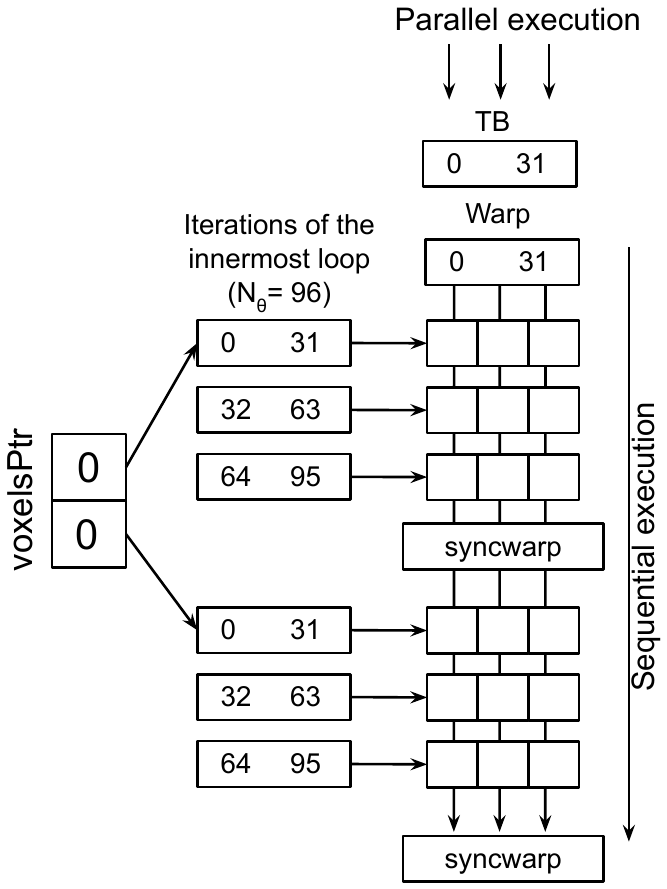} }
    \subfloat[\label{fig:main1d}]{
	    \includegraphics[width=0.5\linewidth,scale=0.5] {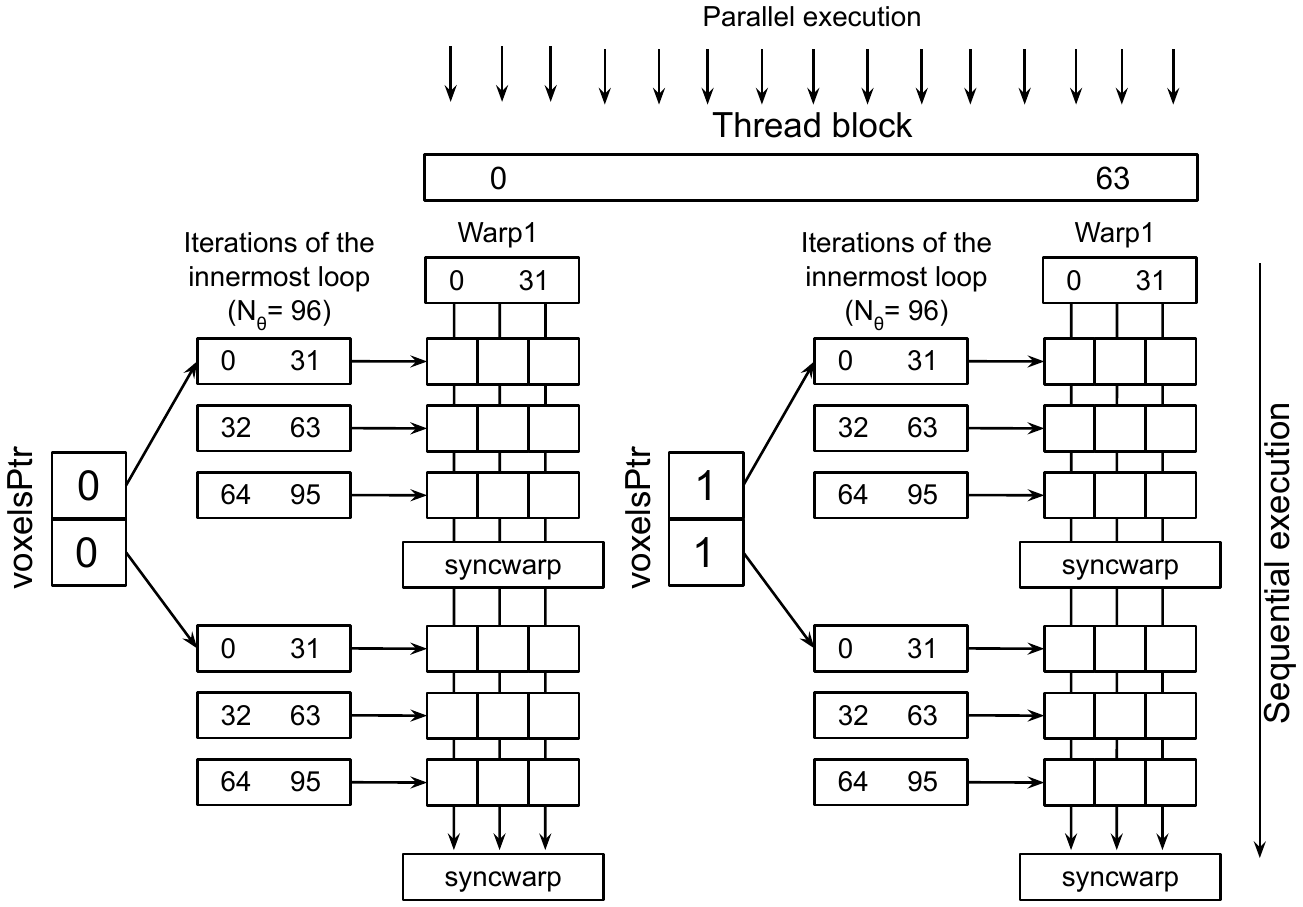} }

    \caption[Various GPU optimizations]{(a) Sequential execution of iterations
	of a sub-vector of the \texttt{voxelsPtr} vector scheduled to single
	thread block. (b) Parallel execution of innermost loop ($N_\theta$=96)
	and sequential execution of a sub-vector of \texttt{voxelsPtr}
	scheduled to three warps and a single thread block. (c) Parallel
	execution of innermost loop ($N_\theta$=96) and sequential execution of
	a sub-vector of the \texttt{voxelsPtr} scheduled to single warp and a
	single thread block. (d) Parallel execution of innermost loop
	($N_\theta$=96) and sequential execution of two distinct sub-vectors of
the \texttt{voxelsPtr} scheduled to single warp and a single thread block.}
\label{fig:main1} \end{figure*} 

\paragraph{\bf Exploiting Fine-grained Parallelism:} \label{subsec:ss} The
reference optimized GPU approach executes the innermost loop of both the SpMV
operations sequentially (Figure~\ref{fig:ref-code}). It was performed due to
the indirect array accesses of the SpMV operations and the concurrent
scheduling of multiple iterations of the outermost loop to a single thread
block; hence, the innermost loop had to be executed sequentially to avoid a
data race. Thus, due to these reasons, the reference GPU approach missed out an
important opportunity to exploit fine-grained parallelism. However, with the
aid of resources and instructions provided by a GPU architecture, we could
exploit fine-grained parallelism; hence, it helps in obtaining substantial
performance improvement in both the SpMV operations. We discuss the different
techniques to achieve fine-grained parallelism in the \texttt{DSC} and
\texttt{WC}.

\noindent \textit{Shared memory:} Shared memory is an on-chip explicitly addressed
memory with significantly lower memory latency than local and global memories
of GPUs. It is key in reducing memory access time when data accessed by the
threads of a thread block exhibit~reuse.

In Figure~\ref{fig:ref-mw}, we notice in the \texttt{DSC} code that the
innermost loop (line~$15$) performing the \texttt{daxpy} operation is executed
sequentially. We used shared memory to execute the iterations of the innermost
loop in parallel, though with the usage of a synchronization barrier. However,
later in Section~\ref{subsec:warp}, we will note that the threads can be
executed without the employment of a memory fence. The added advantage of using
the shared memory is reduced memory bandwidth requirements obtained due to data
reuse of \texttt{YPtr}. Also, note that the size of shared memory required
depends on the diffusion direction~($\bf N_\theta$).

\noindent \textit{Shuffle instruction:} Parallel threads of a thread block share data
using shared memory. However, NVIDIA's Kepler architecture introduced a new
warp-level instruction, named, shuffle instruction
(\texttt{SHFL})~\cite{Demouth2013}, to be utilized when the data is to be
shared directly among the parallel threads of a warp. It leads to a
considerable reduction in latency without the use of shared memory.

In Figure~\ref{fig:ref-mtw}, we observe in the \texttt{WC} code that the
innermost loop (line~$14$) performing \texttt{dot-product} operation is
executed sequentially. The \texttt{dot-product} involves two sub-operations ---
(1) multiply corresponding elements of the vectors, which can be performed in
parallel, (2) perform a reduction, which is performance bottleneck if performed
sequentially.  A popular method to perform reductions in GPUs is to use shared
memory. This method however is dependent on the size of shared memory and
requires the employment of a memory fence, thereby hurting performance. An
alternative method is to use the \texttt{SHFL} instruction~\cite{Demouth2013}.
It helps to share data directly among the parallel threads of a warp, but
requires the usage of a synchronization barrier and shared memory, across the
warps of a thread block. However, later in Section~\ref{subsec:warp}, we will
tackle the synchronization bottleneck as well.  Using \texttt{SHFL}, we
parallelized the \texttt{dot-product} to significantly reduce the execution
time of \texttt{WC}.

Thus, in Figure~\ref{fig:main1b}, we can observe that after incorporating the
fine-grained parallelization, the innermost loop of an SpMV operation is
executed in parallel, where each thread block handles the iterations of a
single sub-vector of \textit{voxelsPtr}. Note that the computations associated
with an iteration of the sub-vector are executed in parallel. However, the
computations across the iterations are executed sequentially, requiring the
\textit{syncthread} barrier in between the iterations. We tried to replace the
\texttt{daxpy} computation in the innermost loop of the \texttt{DSC} code and
the \texttt{dot-product} computation in the innermost loop of the \texttt{WC}
code with appropriate cuBLAS library calls, but were unsuccessful due to the
difficulty in interfacing this from MATLAB.

\paragraph{\bf Reduce Synchronization Overhead by using Warp-based Thread
Execution:} \label{subsec:warp} On NVIDIA GPUs, a warp is a collection of a
certain number of threads (typically $32$) executing the same code in lock-step
and is best used when each thread follows the same execution path. When there
are a number of warps sharing data or performing dependent pieces of
computation, those pieces need to be synchronized and this could impact
performance. As discussed earlier in Section~\ref{subsec:ss}, the SpMV
operations of the LiFE algorithm face a similar challenge.

In Figure~\ref{fig:main1b}, we observe that the iterations of the innermost
loop of the SpMV operations executing in parallel require \textit{syncthread}
barrier across the warps of a thread block. However, by transforming the
innermost loop, multiple warps can be replaced by a single warp.  Note that the
innermost loop parameter depends on $\bf N_\theta$, which is typically a
multiple of $32$ for most of the dMRI datasets ($96$ for dMRI datasets we
used). So the innermost loop is transformed such that the $32$ iterations are
executed in parallel by a warp, and then the next $32$ iterations are executed
in parallel by the same warp, i.e., $\bf N_\theta$/$32$ times sequential
execution (as shown in Figure~\ref{fig:main1c}, $\bf N_\theta$=96 requires
three sequential executions). The advantage of this change is that we can
utilize \textit{syncwarp}, a much less expensive barrier operation when
compared to the \textit{syncthread} barrier. This will also benefit the next
set of optimizations we incorporate to optimize SpMV (discussed later in this
section). However, if $\bf N_\theta$ is not a multiple of $32$ then the zeros
are padded for the \texttt{YPtr} and \texttt{DPtr} vectors to tune their
dimensions to a multiple of $32$. The overhead ($2$-$3$\% of the total
execution time of SBBNNLS) of padding is low considering that it is amortized
across the several iterations of SBBNNLS.

\paragraph{\bf Exploiting Additional Data Reuse:} \label{subsec:multiple}
Earlier in Section~\ref{subsec:tb}, we discussed different ways to partition
computations of the outermost loop of the SpMV operations among the thread
blocks.  We scheduled computations of a single coefficient, voxel or atom
dimension to a single thread block (Figure~\ref{fig:main1a}); so that the
atomic operation hindering the coarse-grained parallelism could be avoided.
Despite this optimization, a thread block could not fully utilize the resources
allocated by a GPU (such as shared memory and cache memory). The reasons for
this were: (1)~the size of $\bf N_\theta$ is small, and (2)~only one warp is
scheduled per thread block because of the optimization discussed in
Section~\ref{subsec:warp} (shown in Figure~\ref{fig:main1c}). 

However, we found that resources allocated for a single thread block could be
utilized optimally (Figure~\ref{fig:main1d}) by scheduling multiple
computations of coefficients, voxels or atoms could to a single thread block.
Thus, this optimization would help to effectively utilize shared memory to
exploit an additional data reuse for the \texttt{YPtr} and \texttt{DPtr}
vectors, thereby leads to reduction of memory bandwidth consumption.
Additionally, the synchronization overhead will also reduce due to the usage of
the \textit{syncwarp} barrier. To obtain near-optimal performance improvements
on this aspect, we empirically determined the right number of computations to
be scheduled for a thread block.  We found that for both the \texttt{DSC} and
\texttt{WC}, \textit{four} computations per thread block provided the
near-optimal performance.

\paragraph{\bf Loop Unrolling:} Loop unrolling is straightforward and well-known
to improve performance by reducing control overhead, providing more instruction
scheduling freedom, and increasing register reuse. Using loop unrolling, we
achieve an additional performance improvement for the $\bf \texttt{DSC}$
operation. However, a similar performance improvement was not observed for the
$\bf \texttt{WC}$ operation because the loop index was static; so the compiler
might have automatically unrolled the loop. We determined the unroll factor by
performing a few experiments and found \textit{eight} was optimal unroll factor
for the $\bf \texttt{DSC}$. We used \texttt{\#pragma~unroll~N} (where N is
unroll factor) to unroll the loop corresponding to the iterations of the
sub-vector of an indirection vector (example \textit{voxelsPtr}) in the CUDA
code of the \texttt{DSC} of SBBNNLS. \\

To summarize the optimization of SpMV on GPUs, first we performed the
target-independent optimizations, followed by the GPU-specific optimizations to
obtain a highly optimized GPU code for the SpMV operations of SBBNNLS.
 
\section{Domain-Specific Language Extensions} \label{sec:dsl} In this section,
we provide a brief overview of the PolyMage DSL and a description of the
constructs we added to the DSL, in order to express sparse matrices and the
related operations used in the LiFE algorithm.  
\begin{figure*}
\hfill
\hfill
\begin{tabular}{l}
\input{code/polymagesparsematvec}\\
\input{code/phinode}
\end{tabular}
\hfill
\hfill
\input{code/sparsematvec}
\caption{(a) PolyMage code for $\bf y=Mw$ operation of LiFE
    (b) PolyMage construct for \texttt{PHI tensor} ($\bf \Phi$) to represent STD-based 
    tensor (c) PolyMage construct for $\bf y=Mw$ operation of LiFE.}
\end{figure*}

\subsection{PolyMage DSL} \cite{mullapudi15asplos}
developed PolyMage, a domain-specific language (DSL) and a compiler for image
processing pipelines. PolyMage automatically generates optimized parallelized
C++ code from a high-level language embedded in the Python. The PolyMage
compiler is based on a polyhedral framework for code transformation and
generation. The constructs used in the PolyMage represents a high-level code in
a polyhedral format. The compiler then performs various optimizations such as
loop fusion, loop tiling across various functions and also marks loop(s)
parallel. Some constructs used in the PolyMage DSL are following:
\texttt{Parameter} construct used to declare a constant value and
\texttt{Variable} construct used to declare a variable which usually serves as
labels for a function dimension. The range of a variable is declared using
\texttt{Interval} construct. \texttt{Function} construct is used to declare a
function mapping from a multi-dimensional integer domain to a scalar value.
\texttt{Conditional} construct is used to specify constraints involving
variables, parameters and function values. \texttt{Case} construct allows a
conditional execution of a computation. We introduce two new constructs to
support sparse matrix and the related operations used in the SBBNNLS algorithm.

\subsection{New Constructs Added to PolyMage} We introduce \texttt{PHI\_Tensor}
construct to represent sparse decomposed tensor to enhance productivity. The
sparse decomposed tensor consists of four vectors: three vectors
\texttt{atomPtr}, \texttt{voxelPtr} and \texttt{fiberPtr} represents the
dimensions of a non-zero value in a connectome tensor and another vector
\texttt{valuesPtr} to represents the actual value of a non-zero index.  We use
\texttt{Matrix} construct already defined in the PolyMage to represent these
four vectors (Figure~\ref{fig:phitensor}).  \textit{Sparse\_matvec} construct
(Figure~\ref{fig:sparsematvec}) is added to perform the sparse matrix-vector
multiplication $\bf y=Mw$ operation (Figure~\ref{fig:DSC}) used in the SBBNNLS
algorithm of the LiFE application.  We obtain a sparse decomposed matrix from
the \texttt{PHI\_Tensor} construct.  Additionally  the dictionary vector
\texttt{DPtr}, the weight vector \texttt{wPtr} and the demeaned diffusion
signal vector \texttt{YPtr} are obtained as a input from the user to update the
\texttt{YPtr} vector. We use the \texttt{Function} construct to execute the
\texttt{Case} construct defined in the function definition based on the $\bf
c1$ and $\bf c2$ \texttt{Condition} construct using the $\bf k$
\texttt{Variable} construct. The high-level PolyMage code used to generate
optimized parallelized C++ code for the sparse matrix operation of the SBBNNLS
algorithm is shown in Figure~\ref{fig:polymagesparsematvec}.  
 
\section{Experimental Evaluation} \label{sec:experimentalevaluation} In this
section, we describe the experimental setup, followed by various code versions
and datasets we evaluated. We then present experimental results while analyzing
them. We show performance improvements we achieved by incorporating the
target-independent and the target-dependent optimizations for SpMV operations
(presented in Section~\ref{sec:optimizations}), then we compare our highly
optimized parallelized CPU implementation and our highly optimized GPU
implementation with the original sequential CPU implementation, a reference
optimized GPU implementation, and ReAl-LiFE's GPU implementation. We also
compare various SpMV code implementations by varying different parameters of
the dMRI datasets.  

\begin{table}[!t]
  \small
  \centering
    \small
  \caption{System details~\label{tab:hardware}}
    \label{tab:systemdetails}
  \begin{tabularx}{0.52\linewidth}{ll}
    \toprule
    Microarchitecture& Intel Skylake \\
    Processors       & {2-socket Intel Xeon Silver 4110} \\
    Clock            & 2.10~GHz                     \\
    Cores            & 16 (8 per socket)                          \\
    Hyperthreading   & disabled                          \\
    Private caches   & 64~KB L1 cache, 1024~KB L2 cache   \\
    Shared cache     & 11,264~KB L3 cache                       \\
    Memory           & 256~GB DDR4 (2.4~GHz) \\
    \midrule
    Microarchitecture (GPU)& NVIDIA Turing \\
    GPU     & NVIDIA GeForce RTX 2080 Ti \\
    Multiprocessors (SMs) & 64\\
    CUDA cores (SPs) & 4352\\
    GPU Base Clock       & 1350 Mhz\\
    L1 cache/shared memory  & 96~KB\\
    L2 cache size  & 5.5~MB\\
    Memory size  & 11.26~GB GDDR6\\
    Memory bandwidth    & 616~GB/s\\
    \midrule
    Matlab version   & 9.5.0.944444 (R2018b) \\
    MRtrix version   & 3.0 \\
    CUDA/NVCC version     & 10.0 \\
    NVCC version     & 10.0.130 \\
    \midrule
    Compiler         & GNU C/C++ (gcc/g++) 6.3.0 \\
    Compiler flags   & -O3 -ptx\\
    OS               & Linux kernel 3.10.0 (64-bit) (CentOS 7)\\
    \bottomrule
  \end{tabularx}
\end{table}

\subsection{Experimental Setup} The evaluation was performed on an NVIDIA
GeForce RTX 2080 Ti GPU and a dual-socket NUMA server with Intel Xeon Silver
4110 processor based on the Intel Skylake architecture. The complete
specification is provided in Table~\ref{tab:systemdetails}. The LiFE
application is originally written in MATLAB with the computationally intensive
SpMV operations of the SBBNNLS algorithm written in C++/CUDA-C++ language.  The
reference optimized code developed by Madhav~\cite{madhav2017}, ReAl-LiFE
implementation~\cite{kumar2019}, and our optimized GPU code are compiled using
NVCC compiler to generate PTX code. The SpMV kernels are represented as a
CUDAKernel object in MATLAB, which is used to invoke the compiled PTX code. The
advantage of using the CUDAKernel object is that the same data is used across
the different iterations of SBBNNLS and need not be transferred back and forth
from the host to device and vice-versa. To compare the execution time of
various tractography algorithms, we use MRtrix~\cite{tournier2012} --- an
advanced tool to analyze the diffusion MRI data. MRtrix generates streamline
tracts for numerous tractography algorithms.

\subsection{Datasets} The evaluation was performed on the
STN96\footnote{https://purl.stanford.edu/rt034xr8593} dMRI dataset collected at
Stanford's Center for Cognitive and Neurobiological Imaging~\cite{stn96data}.

\paragraph*{DS1:} dMRI data was collected at~\cite{stn96data}. The diffusion
signal was measured along the $96$ directions, with the spatial resolution of
$1.5mm$ and the gradient strength of $2000s/mm^2$.  

\paragraph*{DS2:} dMRI data was same as DS1; however, we used MRtrix to
generate streamline tracts in-house for various tractography algorithms such
as: deterministic algorithm (\texttt{Tensor\_DTI}) based on 4-D
diffusion-weighted imaging (DWI)~\cite{Basser2000}, probabilistic algorithm
based 4-D DWI (\texttt{Prob\_DTI})~\cite{jones2008}, fiber assigned continuous
tracking (\texttt{FACT})~\cite{mori1999}, fiber orientation distribution
(\texttt{iFOD1})~\cite{tournier2012}, and spherical deconvolution
(\texttt{SD\_STREAM})~\cite{tournier2012} method.  There are numerous
tractography algorithms available but based on the popularity we choose these
tractography algorithms for our evaluation. 

\subsection{Results and analysis on Multi-core System} In this sub-section, we
present detailed analysis of the target-independent optimizations incorporated
for the SpMV operations running on CPUs, followed by the evaluation of the
CPU-specific optimizations.

\subsubsection{\bf Code Versions:} \label{subsec:cpucodeversion} The various SpMV
code implementations that we use to analyze the performance of the SBBNNLS
algorithm on multi-cores are as follows:

\begin{itemize}

\item \textit{CPU-naive} (Figure~\ref{fig:orig-code}) is the original
	sequential code for the \texttt{DSC} and \texttt{WC} SpMV operations
		developed by Caiafa and Pestilli~\cite{ccaifacode16}.  

\item \textit{CPU-naive-withBLAS} is a variant of \textit{CPU-naive}
	implementation with code fragments replaced by an appropriate BLAS
		call.

\item \textit{CPU-naive-par-withoutBLAS} is a variant of the \textit{CPU-naive}
	version, parallelized by marking the outermost loop parallel though
		having statements with a chance of conflicting data accesses
		marked atomic. Note that the BLAS calls cannot be marked
		atomic. Therefore, the BLAS calls cannot be replaced by the
		code fragment in \textit{Naive-par-withoutBLAS} code version.

\item \textit{CPU-opt} is our highly parallelized optimized C++ code
	implementation with all target-independent optimizations presented in
		Section~\ref{sec:tiopt} and the CPU-specific optimizations
		presented in the Section~\ref{sec:cpuoptimizations}.

\item \textit{CPU-opt-atomic-withoutBLAS} is a variant of \textit{CPU-opt}
	version without usage of a BLAS call and having statements marked
		atomic having a chance of conflicting data dependent accesses.

\item \textit{CPU-opt-withoutBLAS} is a variant of \textit{CPU-opt} version
	without usage of a BLAS call. 

\end{itemize}

\subsubsection{\bf Analysis:} Table~\ref{tab:cpu-dr} shows the execution time in
seconds for \texttt{DSC} and \texttt{WC} operations for different \textit{data
restructuring} methods performed on the \textit{CPU-naive} sequential
implementation. In the table, we observe that the atom-based data restructuring
method is slightly better for both \texttt{DSC} and \texttt{WC} operations.
Also, the execution time of \texttt{DSC} and \texttt{WC} is similar for the
different iterations of SpMV.  Thus, from this table we infer that the
\texttt{DPtr} vector (redirected by the \texttt{atomsPtr}) captures better
reuse compared to the \texttt{YPtr} vector (redirected by the
\texttt{voxelsPtr}).

\begin{table}[!h] \begin{center} \caption{Execution time for \textit{CPU-naive}
	implementation of the SpMV operations for various data restructuring
	methods on Intel Xeon processor.} \label{tab:cpu-dr} \small
	\begin{tabularx}{0.39\linewidth}{ccccc} \toprule
		\multirow{3}[1]{*}{Iterations} & \multicolumn{4}{c}{SpMV
		operation} \\ \cmidrule(lr){2-5} & \multicolumn{2}{c}{DSC}  &
		\multicolumn{2}{c}{WC}\\ \cmidrule(lr){2-3} \cmidrule(lr){4-5}
		& Atom       & Voxel       & Atom      & Voxel   \\ \midrule 1
		& 8.187s     &  8.214s    & 6.320s   & 6.490s \\ 100 & 8.490s
		&  8.228s    & 6.316s   & 6.496s \\ 200 & 8.484s     &  8.227s
		& 6.315s   & 6.500s \\ 300 & 8.465s     &  8.231s    & 6.283s
		& 6.478s\\ 400 & 8.459s     &  8.210s    & 6.286s   & 6.464s \\
	500 & 8.452s     &  8.767s    & 6.301s   & 6.463s \\ \bottomrule
	\end{tabularx} \end{center} \end{table}

\begin{table}[!t] \begin{center} \caption{Execution time of
	\textit{CPU-naive-par-withoutBLAS} implementation of SpMV for different
	\textit{computation partitioning + data restructuring} combinations on
	Intel Xeon processor.} \label{tab:cpu-naive-cs_dr} \small
	\begin{tabularx}{0.80\linewidth}{ccccccc} \toprule &
		\multicolumn{6}{c}{SpMV operation} \\ \cmidrule(lr){2-7}
		Iterations & \multicolumn{2}{c}{DSC}  &
		\multicolumn{4}{c}{WC}\\ \cmidrule(lr){2-3} \cmidrule(lr){4-7}
		& Voxel+Voxel & Coeff+Voxel & Voxel+Voxel & Atom+Atom &
		Coeff+Voxel & Coeff+Atom \\ \midrule 1   & 2.553s & 2.040s  &
		0.905s   & 0.957s   & 0.720s   & 0.678s \\ 100 & 2.663s &
		1.785s  & 0.814s   & 0.904s   & 0.636s   & 0.682s \\ 200 &
		2.451s & 1.778s  & 0.877s   & 0.955s   & 0.640s   & 0.666s \\
		300 & 2.471s & 1.787s  & 0.834s   & 1.001s   & 0.645s   &
		0.673s \\ 400 & 2.412s & 1.778s  & 0.841s   & 0.905s   & 0.646s
		& 0.665s \\ 500 & 2.407s & 1.783s  & 0.811s   & 0.907s   &
		0.660s   & 0.667s \\

    \bottomrule \end{tabularx} \end{center} \end{table}

Table~\ref{tab:cpu-naive-cs_dr} shows the execution time in seconds for
\texttt{DSC} and \texttt{WC} operations for different combinations of
\textit{computations partitioning + data restructuring} methods performed on
\textit{CPU-naive-par-withoutBLAS} implementation (marking the outermost loop
parallel and data dependent statements marked atomic) running on $16$-core
Intel Xeon processor. For \texttt{DSC} operation, we observe that the
\textit{coefficient-based partitioning + voxel-based restructuring} combination
performs better due to efficient usage of the parallelism provided by CPU with
a low load imbalance. In contrast, the \textit{voxel-based partitioning +
voxel-based restructuring} combination does not perform well due to a high load
imbalance. Note that \texttt{DSC} operation involves reduction of \texttt{YPtr}
(with indirection from \texttt{voxelsPtr}); therefore, the voxel-based
technique will capture reuse twice due to read and write access, whereas the
atom-based restructuring will require usage of an atomic operation for the
reduction of the irregularly accessed \texttt{YPtr}. Thus, due to these
reasons we skip the atom-based restructuring for the \texttt{DSC} operation. For
\texttt{WC} operation, we observe that the \textit{coefficient-based
partitioning + atom-based restructuring} combination performs much better
compared to the other combinations. The reason for this is that the
coefficient-based partition exploits the parallelism effectively, on the other
hand the atom-based data restructuring captures the data reuse efficiently. 
Thus, this combination is best for \texttt{WC} operation. Besides this, one can 
observe that the execution time of \texttt{DSC} and \texttt{WC} is similar
for the different iterations of SpMV.

\begin{table}[!t] \centering
    \caption{Execution time of \textit{CPU-opt} implementation of SpMV for
	different \textit{computation partitioning + data restructuring}
	combinations on Intel Xeon processor.} \label{tab:cpu-opt-cs_dr} \small
	\begin{tabularx}{0.80\linewidth}{ccccccc} \toprule &
		\multicolumn{6}{c}{SpMV operation} \\ \cmidrule(lr){2-7}
		Iterations  & \multicolumn{2}{c}{DSC}  &
		\multicolumn{4}{c}{WC}\\ \cmidrule(lr){2-3} \cmidrule(lr){4-7}
		& Voxel+Voxel & Coeff+Voxel & Voxel+Voxel & Atom+Atom &
		Coeff+Voxel & Coeff+Atom \\ \midrule 1   & 0.534s & 0.486s  &
		0.510s   & 0.545s   & 0.482s   & 0.382s \\ 100 & 0.147s &
		0.133s  & 0.496s   & 0.533s   & 0.442s   & 0.379s \\ 200 &
		0.124s & 0.113s  & 0.496s   & 0.524s   & 0.432s   & 0.376s \\
		300 & 0.126s & 0.111s  & 0.503s   & 0.534s   & 0.428s   &
		0.375s \\ 400 & 0.139s & 0.112s  & 0.500s   & 0.538s   & 0.446s
		& 0.375s \\ 500 & 0.122s & 0.111s  & 0.498s   & 0.559s   &
		0.426s   & 0.391s \\

    \bottomrule \end{tabularx}
\end{table}

Table~\ref{tab:cpu-opt-cs_dr} shows the execution time in seconds for
\texttt{DSC} and \texttt{WC} operations performed using \textit{CPU-opt}
implementation (running on $16$-core Intel Xeon processor) for different
\textit{computations partitioning + data restructuring} combinations.  We
observe that for both \texttt{DSC} and \texttt{WC}, the \textit{computation
partitioning + data restructuring} combination that performs best is similar to
that of \textit{CPU-naive-par-withoutBLAS} implementation. However, the
execution time is significantly lower for \textit{CPU-opt} SpMV operations
compared to the \textit{CPU-naive-par-withoutBLAS} implementation due to the
CPU-specific optimizations that we incorporated. Another interesting point to
observe is that the execution time of \texttt{DSC} reduces as the iteration
increases. The reasons for this is due the sparse property of the \texttt{wPtr}
vector. We discuss more about it later in this sub-section.

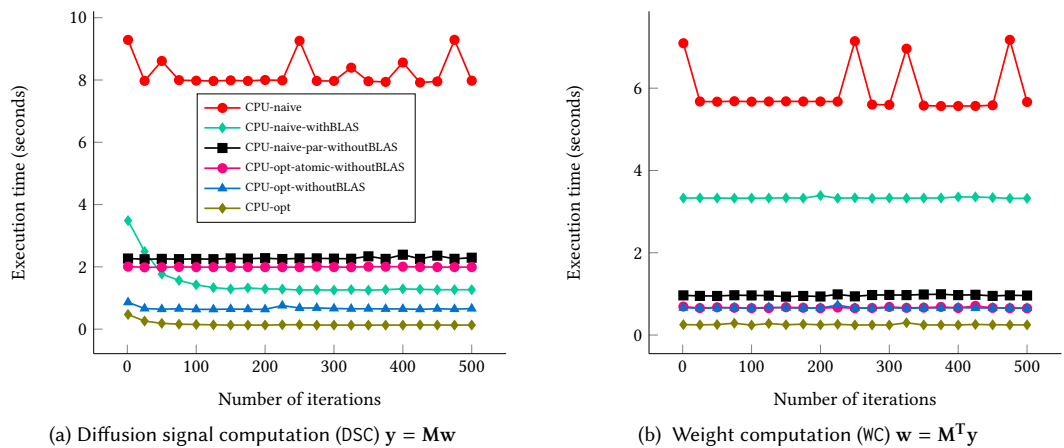
\begin{figure}[!t] \centering \definecolor{trueblue}{rgb}{0.0, 0.45, 0.81}
\definecolor{red}{rgb}{1.0, 0.0, 0.0}
\definecolor{caribbeangreen}{rgb}{0.0, 0.8, 0.6}
\definecolor{black}{rgb}{0.0, 0.0, 0.0}
\definecolor{brightpink}{rgb}{1.0, 0.0, 0.5}
\definecolor{yellow-green}{rgb}{0.6, 0.8, 0.2}
\definecolor{heartgold}{rgb}{0.5, 0.5, 0.0}
\subfloat[Diffusion signal computation (\texttt{DSC}) $\bf y=Mw$ \label{fig:dsc-naive}]{
\begin{tikzpicture}[scale=0.80]
    \begin{axis}[
        legend columns=1,
        legend cell align=left,
        legend style={at={(0.25,0.76)}, anchor=north west, font=\scriptsize,
                     /tikz/column 2/.style={column sep=5pt}},
        ylabel style={align=center}, ylabel = Execution time (seconds),
        xlabel=Number of iterations,
        axis x line*=left,
        axis y line*=left,
        after end axis/.code={},
        every node near coord/.append style={font=\small},
        nodes near coords align={vertical}
        ]
        \addplot[color=red, thick, mark=*] plot coordinates {
        (1,9.2881)
        (25,7.972)
        (50,8.6122)
        (75,7.996)
        (100,7.9779)
        (125,7.9687)
        (150,7.989)
        (175,7.9712)
        (200,8.0009)
        (225,7.988)
        (250,9.2588)
        (275,7.9702)
        (300,7.9708)
        (325,8.3963)
        (350,7.9589)
        (375,7.9388)
        (400,8.5617)
        (425,7.9202)
        (450,7.9536)
        (475,9.2901)
        (500,7.9785)
        };
        \addplot [color=caribbeangreen, thick, mark=diamond*] plot coordinates {
        (1,3.4863)
        (25,2.4925)
        (50,1.7687)
        (75,1.566)
        (100,1.4244)
        (125,1.3332)
        (150,1.2902)
        (175,1.3236)
        (200,1.2923)
        (225,1.2852)
        (250,1.256)
        (275,1.2637)
        (300,1.2509)
        (325,1.2685)
        (350,1.2499)
        (375,1.2691)
        (400,1.2923)
        (425,1.2796)
        (450,1.2658)
        (475,1.2675)
        (500,1.2665)
        };
        \addplot [color=black, thick, mark=square*] plot coordinates {
        (1,2.2677)
        (25,2.2444)
        (50,2.2563)
        (75,2.2478)
        (100,2.255)
        (125,2.2472)
        (150,2.272)
        (175,2.2632)
        (200,2.2794)
        (225,2.2518)
        (250,2.2727)
        (275,2.2731)
        (300,2.2655)
        (325,2.261)
        (350,2.3378)
        (375,2.256)
        (400,2.3905)
        (425,2.2574)
        (450,2.359)
        (475,2.2579)
        (500,2.2935)
        };
        \addplot [color=brightpink, thick, mark=*] plot coordinates {
        (1,2.0085)
        (25,1.9858)
        (50,1.9879)
        (75,1.9994)
        (100,1.991)
        (125,1.9913)
        (150,1.9944)
        (175,1.9934)
        (200,1.9915)
        (225,1.9895)
        (250,1.9908)
        (275,2.0134)
        (300,1.9926)
        (325,1.9898)
        (350,2.0071)
        (375,2.0049)
        (400,2.0068)
        (425,1.995)
        (450,1.9932)
        (475,1.9915)
        (500,1.9925)
        };
        \addplot [color=trueblue, thick, mark=triangle*] plot coordinates {
        (1,0.86146)
        (25,0.66215)
        (50,0.64053)
        (75,0.65364)
        (100,0.63551)
        (125,0.6352)
        (150,0.64507)
        (175,0.63951)
        (200,0.63596)
        (225,0.75648)
        (250,0.67889)
        (275,0.68155)
        (300,0.66434)
        (325,0.65207)
        (350,0.65584)
        (375,0.64972)
        (400,0.64894)
        (425,0.63812)
        (450,0.65311)
        (475,0.64271)
        (500,0.65904)
        };
        \addplot [color=heartgold, thick, mark=diamond*] plot coordinates {
        (1,0.4714)
        (25,0.26103)
        (50,0.18629)
        (75,0.15992)
        (100,0.14936)
        (125,0.13814)
        (150,0.13153)
        (175,0.13177)
        (200,0.13102)
        (225,0.13999)
        (250,0.14306)
        (275,0.13029)
        (300,0.13134)
        (325,0.12926)
        (350,0.12932)
        (375,0.12956)
        (400,0.12964)
        (425,0.13707)
        (450,0.1316)
        (475,0.13053)
        (500,0.13211)
        };
        \addlegendentry{CPU-naive}
        \addlegendentry{CPU-naive-withBLAS}
        \addlegendentry{CPU-naive-par-withoutBLAS}
        \addlegendentry{CPU-opt-atomic-withoutBLAS}
        \addlegendentry{CPU-opt-withoutBLAS}
        \addlegendentry{CPU-opt}
    \end{axis}
 \end{tikzpicture}
} \definecolor{trueblue}{rgb}{0.0, 0.45, 0.81}
\definecolor{red}{rgb}{1.0, 0.0, 0.0}
\definecolor{caribbeangreen}{rgb}{0.0, 0.8, 0.6}
\definecolor{black}{rgb}{0.0, 0.0, 0.0}
\definecolor{brightpink}{rgb}{1.0, 0.0, 0.5}
\definecolor{yellow-green}{rgb}{0.6, 0.8, 0.2}
\definecolor{heartgold}{rgb}{0.5, 0.5, 0.0}
\subfloat[ Weight computation (\texttt{WC}) $\bf w=M^Ty$ \label{fig:wc-naive}]{
\begin{tikzpicture}[scale=0.80]
    \begin{axis}[
        legend columns=1,
        legend cell align=left,
        legend style={at={(0.35,0.70)}, anchor=north west, font=\footnotesize,
                     /tikz/column 2/.style={column sep=5pt}},
        ylabel style={align=center}, ylabel = Execution time (seconds),
        xlabel=Number of iterations,
        axis x line*=left,
        axis y line*=left,
        after end axis/.code={},
        every node near coord/.append style={font=\scriptsize},
        nodes near coords align={vertical}
        ]
        \addplot[color=red, thick, mark=*] plot coordinates {
        (1,7.0925)
        (25,5.6763)
        (50,5.6713)
        (75,5.6823)
        (100,5.6726)
        (125,5.6759)
        (150,5.6799)
        (175,5.6768)
        (200,5.6762)
        (225,5.6749)
        (250,7.1419)
        (275,5.6012)
        (300,5.5954)
        (325,6.9599)
        (350,5.576)
        (375,5.5656)
        (400,5.5659)
        (425,5.5675)
        (450,5.584)
        (475,7.1746)
        (500,5.665)
        };
        \addplot [color=caribbeangreen, thick, mark=diamond*] plot coordinates {
        (1,3.3287)
        (25,3.3309)
        (50,3.329)
        (75,3.3262)
        (100,3.3276)
        (125,3.3284)
        (150,3.3353)
        (175,3.3275)
        (200,3.3899)
        (225,3.3248)
        (250,3.3348)
        (275,3.3238)
        (300,3.3302)
        (325,3.3268)
        (350,3.3291)
        (375,3.3322)
        (400,3.3557)
        (425,3.3558)
        (450,3.3402)
        (475,3.3212)
        (500,3.3222)
        };
        \addplot [color=black, thick, mark=square*] plot coordinates {
        (1,0.96428)
        (25,0.9515)
        (50,0.94938)
        (75,0.96744)
        (100,0.96292)
        (125,0.95788)
        (150,0.9339)
        (175,0.94952)
        (200,0.93691)
        (225,0.98782)
        (250,0.94082)
        (275,0.97273)
        (300,0.9744)
        (325,0.97102)
        (350,0.98547)
        (375,0.99118)
        (400,0.97034)
        (425,0.9784)
        (450,0.95089)
        (475,0.96544)
        (500,0.96039)
        };
        \addplot [color=brightpink, thick, mark=*] plot coordinates {
        (1,0.69204)
        (25,0.65615)
        (50,0.67307)
        (75,0.66282)
        (100,0.65273)
        (125,0.65521)
        (150,0.67598)
        (175,0.66327)
        (200,0.65188)
        (225,0.67073)
        (250,0.65229)
        (275,0.65778)
        (300,0.68358)
        (325,0.66102)
        (350,0.66625)
        (375,0.68359)
        (400,0.65665)
        (425,0.70949)
        (450,0.66024)
        (475,0.65206)
        (500,0.64992)
        };
        \addplot [color=trueblue, thick, mark=triangle*] plot coordinates {
        (1,0.66364)
        (25,0.65305)
        (50,0.6583)
        (75,0.6616)
        (100,0.65199)
        (125,0.67377)
        (150,0.6635)
        (175,0.65187)
        (200,0.65585)
        (225,0.72742)
        (250,0.65716)
        (275,0.65211)
        (300,0.66456)
        (325,0.65064)
        (350,0.65893)
        (375,0.65865)
        (400,0.66825)
        (425,0.65485)
        (450,0.65782)
        (475,0.65767)
        (500,0.65373)
        };
        \addplot [color=heartgold, thick, mark=diamond*] plot coordinates {
        (1,0.25324)
        (25,0.24774)
        (50,0.25482)
        (75,0.28676)
        (100,0.241)
        (125,0.28087)
        (150,0.25021)
        (175,0.26386)
        (200,0.24907)
        (225,0.26103)
        (250,0.24384)
        (275,0.24565)
        (300,0.24444)
        (325,0.29834)
        (350,0.245)
        (375,0.2461)
        (400,0.24476)
        (425,0.2585)
        (450,0.24904)
        (475,0.24737)
        (500,0.24678)
        };
    \end{axis}
 \end{tikzpicture}
}
\caption{Execution time (in seconds) of SpMV used in the LiFE with various
optimizations running on Intel Xeon processor.} \label{fig:cpu-optimization}
\end{figure}

Figure~\ref{fig:cpu-optimization} reports absolute execution time in seconds
for different CPU code implementations of SpMV ($\bf Mw$ and $\bf M^Ty$) for
different iterations of the SBBNNLS algorithm. We observe that by marking the
outermost loop parallel in the \textit{Naive-par-withoutBLAS} version achieved
a speedup of $4.3\times$ over the \textit{CPU-naive} version. However, it did
not improve the performance significantly due to following reasons: (a) poor data locality
captured for \texttt{YPtr} and \texttt{DPtr} vectors, and (b) statements marked
atomic due to a chance of conflicting data accesses. We also notice that the
\textit{CPU-opt-atomic-withoutBLAS} version shows comparable performance with
the \textit{Naive-par-withoutBLAS} version for the same reasons (the statements
were marked atomic), though slightly better due to the improved data reuse. For
the \texttt{DSC} operation, we calculated the average execution time over the
$500$ iterations of SBBNNLS and obtained a speedup of $12.43\times$ for
\textit{CPU-opt} version over \textit{CPU-opt-atomic-withoutBLAS} version.
However, after incorporating the target-independent optimizations and the
efficient synchronization-free thread mapping optimization, we not only
obtained better data reuse but were also able to mark the outermost loop
parallel without the usage an atomic operation (for the \texttt{DSC}
operation). Overall, for complete execution of SBBNNLS we obtained a speedup of
$27.25\times$ and $6.33\times$ for \textit{CPU-opt} version over the
\textit{CPU-naive} and \textit{CPU-naive-par-withoutBLAS} respectively. Later
in this sub-section, we will discuss more about benefit of code parallelization
of the SpMV operations of LiFE.

We also observe that mapping to a BLAS call significantly improved the
performance of both the \textit{CPU-naive} and the \textit{CPU-opt} versions of
the \texttt{DSC} and \texttt{WC} computations. We notice that for the
\texttt{DSC} operation, as the number of iteration increases, the execution
time reduces remarkably and becomes stable thereafter; the reason for this
improvement is the weight vector (\texttt{wPtr}) becomes sparser. So when the
vector \texttt{wPtr} is used as a scalar in the argument of the
(\texttt{daxpy}) BLAS call, the invocation of the call is evaded to avoid
unnecessary computations. We computed the average execution time of the
\texttt{DSC} operations over $500$ iterations and obtained a speedup of
$5.5\times$ for \textit{CPU-naive-withBLAS} version over the \textit{CPU-naive}
version. Similarly, we achieved a speedup of $4.81\times$ for the
\textit{CPU-opt} version over the \textit{CPU-opt-withoutBLAS} version. Note
that in the \texttt{WC} operation, the similar performance improvement was not
observed due the set of computations it involved.

\begin{table}[!t] \caption{Total execution time (in min) up till different
    iterations of the SBBNNLS algorithm for a different number of cores on
    Intel Xeon processor. The baseline is \textit{CPU-naive} version.}
    \label{tab:thread}
    \small
    \begin{tabularx}{1\linewidth}{lcXXXXXXXXXXXX}
        \toprule
        Code        &Iters    &\multicolumn{6}{c}{Execution time (in min)}    &\multicolumn{6}{c}{Speedup} \\ \cmidrule(lr){3-8} \cmidrule(lr){9-14} 
                    &  &  & 2    &  4   &  8   & 12   & 16    &    1 &   2  &   4  &  8   & 12   &  16 \\\midrule
        CPU-naive   & 10  & 5.24 & 5.73 & 3.14 & 1.85 & 1.37  & 1.11 & 1.00 & 0.91 & 1.66 & 2.82 & 3.81 & 4.73 \\
                    & 100 & 45.1 & 54.5 & 30.1 & 17.9 & 13.1  & 10.5 & 1.00 & 0.82 & 1.49 & 2.51 & 3.42 & 4.29 \\
                    & 500 & 225  & 271  & 149  & 88.6 & 65.6  & 52.2 & 1.00 & 0.82 & 1.51 & 2.53 & 3.42 & 4.30 \\ \midrule
        CPU-opt     & 10  & 1.91 & 1.22 & 0.66 & 0.41 & 0.35  & 0.31 & 2.74 & 4.29 & 7.86 & 12.5 & 14.9 & 17.2 \\
                    & 100 & 14.3 & 8.86 & 4.76 & 2.78 & 2.22  & 1.97 & 3.13 & 5.08 & 9.46 & 16.1 & 20.3 & 22.8 \\
                    & 500 & 61.8 & 39.1 & 20.8 & 12.1 & 9.43  & 8.29 & 3.63 & 5.74 & 10.8 & 18.5 & 23.8 & 27.2 \\
        \bottomrule
    \end{tabularx}
\end{table}

Table~\ref{tab:thread} shows the total execution time up till different
iterations of the SBBNNLS algorithm and speedups achieved with different number
of threads. We compare the performance of the \textit{CPU-naive} version (also
used as base version) with \textit{CPU-naive-par-withoutBLAS} version and the
\textit{CPU-opt} version. We observe that for the
\textit{Naive-par-withoutBLAS} version, the speedup remains similar for the
different iterations. In addition to that, as the number of threads increases
the performance does not scale well. However, for the \textit{CPU-opt} code
version the performance improves for different iterations of the SBBNNLS
algorithm.  Also it is worth noting that the performance scales well up till
$8$ threads due to improved data reuse, but because of NUMA effects does not
scale further. As discussed earlier in this sub-section, mapping a code
fragment of the \texttt{DSC} operation to the BLAS call leverages the sparse
nature of the \texttt{wPtr} vector. From Table~\ref{tab:thread} the same can be
seen, the more the number of iteration the better is the speedup achieved for
the \textit{CPU-opt} version. The \textit{CPU-naive} and
\textit{Naive-par-withoutBLAS} code versions does not use a BLAS call; hence,
the speedup remains the same for them due to the execution of the unnecessary
computations.

Summarizing the results for CPUs, for the \texttt{DSC} operation we achieve
optimal performance by incorporating the voxel-based data restructuring
technique. For the \texttt{WC} operation, we achieve optimum performance by
incorporating the atom-based data restructuring technique. Once the data was
restructured, optimizations such as loop tiling and code parallelism helped
obtaining coarse-grained parallelism. We achieved significantly better
performance improvement by mapping to BLAS calls for exploiting fine-grained
parallelism.

\subsection{Results and analysis on GPU} In this sub-section, we present detailed
analysis of the target-independent optimizations incorporated for the SpMV
operations running on GPUs, followed by the evaluation of the GPU-specific
optimizations.

\subsubsection{\bf Code Versions:} The various SpMV code implementations that we use to
analyze the performance of the SBBNNLS algorithm on GPUs are as follows:
\begin{itemize} 

\item \textit{Ref-opt} is a reference optimized GPU code developed by
	Madhav~\cite{madhav2017}, on a similar set of CPU optimization
		mentioned for the \textit{CPU-opt} implementation. For the SpMV
		operations, the \textit{Ref-opt} code reorders the data based
		on the atom dimension to exploit data reuse and also uses the
		coefficient-based partitioning to achieve
		coarse-grained~parallelism.

\item \textit{ReAl-LiFE} is a GPU-accelerate implementation using the
	voxel-based data restructuring and the voxel-based computation
		partitioning for both \texttt{DSC} and \texttt{WC} operations.
		In addition, the ReAl-LiFE implementations uses shared memory
		for \texttt{DSC} and shared memory + shuffle instruction for
		\texttt{WC} operations to achieve fine-grained parallelism with
		single-warp based execution.

\item \textit{GPU-opt} is our optimized GPU code implementation with all the
	optimizations mentioned in Section~\ref{sec:gpuoptimizations}. In
		contrast to ReAl-LiFE implementation, we added following
		optimizations: (1)~automated selection of the data
		restructuring + computation partitioning combination at
		run-time, (2)~utilized only shuffle instruction to exploit
		fine-grained parallelism for \texttt{WC}, (3)~scheduled
		multiple computations to a thread block, and (4)~exploited the
		sparse property of the \texttt{wPtr} vector.

\end{itemize} 

\subsubsection{\bf Analysis:} \label{sec:gpuanalysis} 
\begin{table}[!t]
\begin{center}
    \caption{Execution time of \textit{Ref-opt} implementation of the SpMV 
    operations for various data restructuring techniques on NVIDIA GPU.}
    \label{tab:gpu-ds}
    \small
    \begin{tabularx}{0.39\linewidth}{ccccc}
    \toprule
    \multirow{3}[1]{*}{Iterations} & \multicolumn{4}{c}{SpMV operation} \\ \cmidrule(lr){2-5}
        & \multicolumn{2}{c}{DSC}  & \multicolumn{2}{c}{WC}\\ \cmidrule(lr){2-3} \cmidrule(lr){4-5}
        & Atom       & Voxel       & Atom      & Voxel   \\ \midrule
    1   & 1.025s     &  2.087s     & 0.310s     & 0.311s \\
    100 & 0.185s     &  0.219s     & 0.316s     & 0.320s \\
    200 & 0.166s     &  0.190s     & 0.319s     & 0.320s \\
    300 & 0.162s     &  0.187s     & 0.319s     & 0.320s\\
    400 & 0.162s     &  0.186s     & 0.319s     & 0.320s \\
    500 & 0.162s     &  0.186s     & 0.319s     & 0.320s \\
    \bottomrule
    \end{tabularx}
\end{center}
\end{table}

 Table~\ref{tab:gpu-ds} reports
the execution time in seconds for \texttt{DSC} and \texttt{WC} operations at
different iterations of SBBNNLS for various \textit{data restructuring}
techniques discussed in Section~\ref{subsec:ds}.  Evaluation was performed on
the \textit{Ref-opt + data-restructure} code --- a modification of the
\textit{Ref-opt} GPU code obtained by incorporating the data restructuring
optimization. We observe that the performance of the atom-based data
restructuring is surprisingly better than the voxel-based data restructuring
for the \texttt{DSC} computation. The reason for this is that the voxel-based
approach achieve good data reuse; however, due to the usage of an atomic
operation the overhead is high.  Though, later in this sub-section, we will
discern that when other optimizations are incorporated, the voxel-based data
restructuring technique outruns the atom-based technique.  In the case of
\texttt{WC}, we observe that the atom-based and voxel-based restructuring
techniques achieve a similar order of performance because the data reuse is
obtained either for the \texttt{YPtr} vector or the \texttt{DPtr} vector.

\begin{table}[!t]
\begin{center}
    \caption{Execution time of \textit{Ref-opt} implementation of the SpMV
    operations for different \textit{computation partitioning + data
    restructuring} combinations on NVIDIA GPU.} 
    \label{tab:ref-opt-cs_dr}
    \small
    \begin{tabularx}{0.80\linewidth}{ccccccc}
    \toprule
                & \multicolumn{6}{c}{SpMV operation} \\ \cmidrule(lr){2-7}
    Iter(s)  & \multicolumn{2}{c}{DSC}  & \multicolumn{4}{c}{WC}\\ \cmidrule(lr){2-3} \cmidrule(lr){4-7}
                & Voxel+Voxel & Coeff+Voxel & Voxel+Voxel   & Atom+Atom & Coeff+Voxel   & Coeff+Atom \\ \midrule
            1   & 0.318s      & 1.025s      & 0.311s        & 0.313s    & 0.188s        & 0.122s \\
            100 & 0.057s      & 0.185s      & 0.318s        & 0.320s    & 0.184s        & 0.121s \\
            200 & 0.053s      & 0.166s      & 0.321s        & 0.320s    & 0.184s        & 0.121s \\
            300 & 0.052s      & 0.162s      & 0.321s        & 0.320s    & 0.184s        & 0.121s \\
            400 & 0.052s      & 0.162s      & 0.321s        & 0.320s    & 0.182s        & 0.121s \\
            500 & 0.052s      & 0.162s      & 0.322s        & 0.320s    & 0.184s        & 0.120s \\

    \bottomrule
    \end{tabularx}
\end{center}
\end{table}

Table~\ref{tab:ref-opt-cs_dr} shows the execution time in seconds for
\texttt{DSC} and \texttt{WC} operations performed using \textit{Ref-opt}
implementation for different combinations of \textit{computations partitioning
+ data restructuring} methods. For \texttt{DSC} operation, we observe that the
\textit{voxel-based partitioning + voxel-based restructuring} combination
performs better compared to the \textit{coefficient-based partitioning +
voxel-based restructuring}.  As discussed earlier in Section~\ref{subsec:tb},
the reason for this is that the load imbalance issue on GPUs caused due to
partitioning based on voxel dimension is low considering its massive
parallelism. Additionally, the number of iterations of the outermost loop ($\bf
N_c$) is much larger than maximum possible thread blocks that can be scheduled
to a GPU. Hence, this combination performs good for \texttt{DSC} operation.  In
contrast to that, the coefficient-based partitioning performs poorly because of
the reduction of the \texttt{YPtr} has dependent accesses at runtime;
therefore, this partitioning method have a high synchronization overhead due to
the usage of an atomic operation to avoid data races. For \texttt{WC}
operation, the combination of \textit{coefficient-based partitioning +
atom-based restructuring} performs best compared to others. The reason for this
is that the coefficient-based partitioning exploits parallelism of GPUs
effectively, on the other hand atom-based data restructuring leverages data
reuse efficiently. Also, one can observe that the execution time for both
\texttt{DSC} and \texttt{WC} operations are same for different iterations of
SBBNNLS; therefore, the sparse property of \texttt{wPtr} is not exploited
efficiently by different combinations of \textit{computations partitioning +
data restructuring} methods in \texttt{Ref-opt} implementation.

\begin{table}[!t]
\begin{center}
    \caption{Execution time of the \textit{GPU-opt} implementation of the SpMV
    operations for different \textit{computation partitioning + data
    restructuring} combinations on NVIDIA GPU.} \label{tab:gpu-opt-cs_dr}
    \small
    \begin{tabularx}{0.80\linewidth}{ccccccc}
    \toprule
                & \multicolumn{6}{c}{SpMV operation} \\ \cmidrule(lr){2-7}
    Iter(s)  & \multicolumn{2}{c}{DSC}  & \multicolumn{4}{c}{WC}\\ \cmidrule(lr){2-3} \cmidrule(lr){4-7}
                & Voxel+Voxel & Coeff+Voxel & Voxel+Voxel & Atom+Atom & Coeff+Voxel & Coeff+Atom \\ \midrule
            1   & 0.041s & 2.431s  & 0.074s   & 0.069s   & 0.049s   & 0.057s \\
            100 & 0.017s & 0.141s  & 0.064s   & 0.065s   & 0.047s   & 0.044s \\
            200 & 0.015s & 0.094s  & 0.064s   & 0.064s   & 0.047s   & 0.044s \\
            300 & 0.015s & 0.089s  & 0.064s   & 0.065s   & 0.047s   & 0.044s \\
            400 & 0.015s & 0.089s  & 0.064s   & 0.065s   & 0.047s   & 0.044s \\
            500 & 0.015s & 0.089s  & 0.065s   & 0.065s   & 0.047s   & 0.044s \\

    \bottomrule
    \end{tabularx}
\end{center}
\end{table}

Table~\ref{tab:gpu-opt-cs_dr} shows the execution time in seconds for
\texttt{DSC} and \texttt{WC} operations performed using \textit{GPU-opt}
implementation for different combinations of \textit{computations partitioning
+ data restructuring} methods. We observe that for both \texttt{DSC} and
\texttt{WC}, the \textit{computation partitioning + data restructuring}
combination that performs best is similar to that of \textit{Ref-opt}
implementation. However, the execution time is significantly lower for
\textit{GPU-opt} compared to \textit{Ref-opt} implementation due to the
GPU-specific optimizations we incorporated. Additionally, one can observe that
the execution time of \texttt{DSC} reduces as the iteration increases due to
the sparse property of \texttt{wPtr} vector (discussed in
Section~\ref{subsec:matrix-computation}).
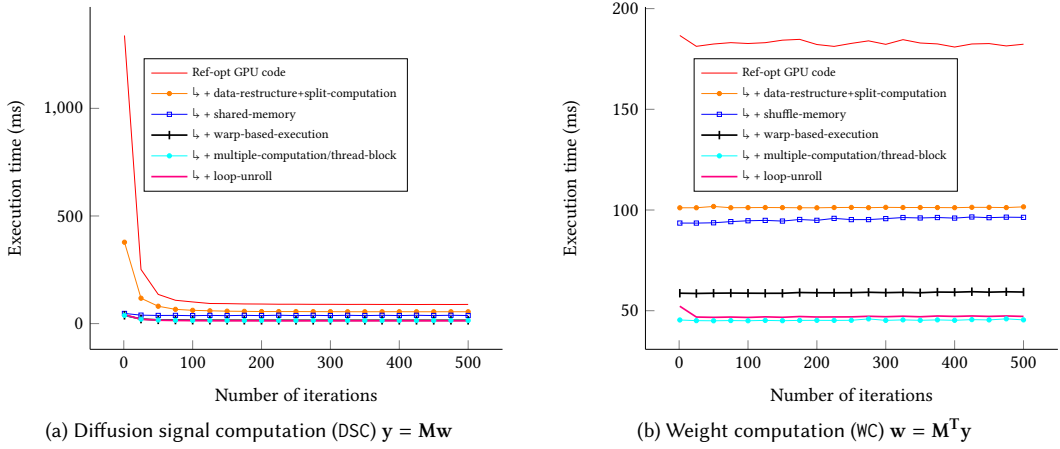
\begin{figure}[!t] 
\centering 
\definecolor{trueblue}{rgb}{0.0, 0.45, 0.81}
\definecolor{red}{rgb}{1.0, 0.0, 0.0}
\definecolor{caribbeangreen}{rgb}{0.0, 0.8, 0.6}
\definecolor{black}{rgb}{0.0, 0.0, 0.0}
\definecolor{brightpink}{rgb}{1.0, 0.0, 0.5}
\definecolor{yellow-green}{rgb}{0.6, 0.8, 0.2}
\definecolor{heartgold}{rgb}{0.5, 0.5, 0.0}
\definecolor{darkgreen}{rgb}{0.0, 0.2, 0.13}
\definecolor{cyan}{rgb}{0.0, 1.0, 1.0}

\subfloat[Diffusion signal computation (\texttt{DSC}) $\bf y=Mw$ \label{fig:dsc}]{
\begin{tikzpicture}[scale=0.80]
    \begin{axis}[
        legend columns=1,
        legend cell align=left,
        legend style={at={(0.13,0.85)}, anchor=north west, font=\scriptsize,
                     /tikz/column 2/.style={column sep=5pt}},
        ylabel style={align=center}, ylabel = Execution time (ms),
        xlabel=Number of iterations,
        axis x line*=left,
        axis y line*=left,
        after end axis/.code={},
        every node near coord/.append style={font=\scriptsize},
        nodes near coords align={vertical}
        ]
         \addplot[color=red, thin, mark=none] plot coordinates {
        (1  ,1338.5)
        (25 ,0251.54)
        (50 ,0135.67)
        (75 ,0108.43)
        (100,0100.63)
        (125,0093.42)
        (150,0092.801)
        (175,0091.326)
        (200,0090.908)
        (225,0089.982)
        (250,0090.075)
        (275,0089.517)
        (300,0089.601)
        (325,0089.181)
        (350,0089.369)
        (375,0088.97)
        (400,0089.173)
        (425,0088.693)
        (450,0089.06)
        (475,0088.774)
        (500,0089.131)
        };
        \addplot [color=orange, thin, mark=*, mark size=1pt] plot coordinates {
        (1,  378.11)
        (25, 117.93)
        (50, 080.673)
        (75, 066.298)
        (100,061.712)
        (125,059.493)
        (150,057.704)
        (175,056.941)
        (200,056.002)
        (225,055.965)
        (250,055.902)
        (275,055.596)
        (300,055.39)
        (325,055.103)
        (350,055.026)
        (375,055.138)
        (400,055.238)
        (425,054.976)
        (450,054.956)
        (475,055.111)
        (500,054.997)
        };
        \addplot [color=blue, thin, mark=square, mark size=1pt] plot coordinates {
        (1,  47.169)
        (25, 39.279)
        (50, 37.84)
        (75, 38.099)
        (100,36.853)
        (125,38.165)
        (150,36.954)
        (175,38.317)
        (200,37.119)
        (225,38.899)
        (250,37.181)
        (275,38.346)
        (300,37.161)
        (325,38.525)
        (350,37.243)
        (375,38.504)
        (400,37.373)
        (425,38.638)
        (450,37.388)
        (475,38.674)
        (500,37.43)
        };
        \addplot [color=black, thick, mark=|, mark size=2pt] plot coordinates {
        (1,  38.443)
        (25, 21.019)
        (50, 17.318)
        (75, 16.421)
        (100,15.712)
        (125,15.541)
        (150,15.325)
        (175,15.339)
        (200,15.263)
        (225,15.242)
        (250,15.153)
        (275,15.227)
        (300,15.149)
        (325,15.174)
        (350,15.158)
        (375,15.282)
        (400,15.15)
        (425,15.2)
        (450,15.165)
        (475,15.231)
        (500,15.208)
        };
        \addplot [color=cyan, thin, mark=*, mark size=1pt] plot coordinates {
        (1,  38.248)
        (25, 20.432)
        (50, 16.647)
        (75, 15.331)
        (100,14.831)
        (125,14.633)
        (150,14.435)
        (175,14.396)
        (200,14.498)
        (225,14.322)
        (250,14.413)
        (275,14.861)
        (300,14.173)
        (325,14.271)
        (350,14.209)
        (375,14.264)
        (400,14.21)
        (425,14.259)
        (450,14.221)
        (475,14.793)
        (500,14.434)
        };
        \addplot [color=brightpink, thick, mark=none, mark size=1.2pt] plot coordinates {
        (1,  41.714)
        (25, 20.067)
        (50, 15.747)
        (75, 14.329)
        (100,13.786)
        (125,13.574)
        (150,13.344)
        (175,13.293)
        (200,13.15)
        (225,13.141)
        (250,13.079)
        (275,13.131)
        (300,13.09)
        (325,13.112)
        (350,13.078)
        (375,13.136)
        (400,13.13)
        (425,13.099)
        (450,13.089)
        (475,13.129)
        (500,13.097)
        };
        \addlegendentry{Ref-opt GPU code}
        \addlegendentry{\rotatebox[origin=c]{180}{$\Lsh$} + data-restructure+split-computation}
        \addlegendentry{\rotatebox[origin=c]{180}{$\Lsh$} + shared-memory}
        \addlegendentry{\rotatebox[origin=c]{180}{$\Lsh$} + warp-based-execution}
        \addlegendentry{\rotatebox[origin=c]{180}{$\Lsh$} + multiple-computation/thread-block}
        \addlegendentry{\rotatebox[origin=c]{180}{$\Lsh$} + loop-unroll}
    \end{axis}
 \end{tikzpicture}
} \definecolor{trueblue}{rgb}{0.0, 0.45, 0.81}
\definecolor{red}{rgb}{1.0, 0.0, 0.0}
\definecolor{caribbeangreen}{rgb}{0.0, 0.8, 0.6}
\definecolor{black}{rgb}{0.0, 0.0, 0.0}
\definecolor{brightpink}{rgb}{1.0, 0.0, 0.5}
\definecolor{yellow-green}{rgb}{0.6, 0.8, 0.2}
\definecolor{heartgold}{rgb}{0.5, 0.5, 0.0}
\subfloat[Weight computation (\texttt{WC}) $\bf w=M^Ty$ \label{fig:wc}]{
\begin{tikzpicture}[scale=0.80]
    \begin{axis}[
        legend columns=1,
        legend cell align=left,
        legend style={at={(0.117,0.85)}, anchor=north west, font=\scriptsize,
                     /tikz/column 2/.style={column sep=5pt}},
        ylabel style={align=center}, ylabel = Execution time (ms),
        xlabel=Number of iterations,
        axis x line*=left,
        axis y line*=left,
        after end axis/.code={},
        every node near coord/.append style={font=\scriptsize},
        nodes near coords align={vertical}
        ]
        \addplot[color=red, thin, mark=none] plot coordinates {
        (1,  186.65)
        (25, 181.27)
        (50, 182.42)
        (75, 183.1)
        (100,182.67)
        (125,183.08)
        (150,184.33)
        (175,184.73)
        (200,182.17)
        (225,181.24)
        (250,182.73)
        (275,183.99)
        (300,182.26)
        (325,184.58)
        (350,182.92)
        (375,182.47)
        (400,180.92)
        (425,182.44)
        (450,182.64)
        (475,181.47)
        (500,182.32)
        };
        \addplot [color=orange, thin, mark=*, mark size=1pt] plot coordinates {
        (1,  101.1)
        (25, 101.12)
        (50, 101.77)
        (75, 101.13)
        (100,101.19)
        (125,101.24)
        (150,101.19)
        (175,101.12)
        (200,101.1)
        (225,101.23)
        (250,101.28)
        (275,101.16)
        (300,101.32)
        (325,101.24)
        (350,101.24)
        (375,101.21)
        (400,101.16)
        (425,101.33)
        (450,101.31)
        (475,101.21)
        (500,101.57)
        };
        \addplot [color=blue, thin, mark=square, mark size=1pt] plot coordinates {
        (1,  93.517)
        (25, 93.503)
        (50, 93.657)
        (75, 94.231)
        (100,94.659)
        (125,94.846)
        (150,94.511)
        (175,95.285)
        (200,94.875)
        (225,95.842)
        (250,95.232)
        (275,95.243)
        (300,95.722)
        (325,96.256)
        (350,96.049)
        (375,96.27)
        (400,95.985)
        (425,96.537)
        (450,96.2)
        (475,96.444)
        (500,96.343)
        };
        
        \addplot [color=black, thick, mark=|, mark size=2pt] plot coordinates {
        (1,  58.689)
        (25, 58.565)
        (50, 58.697)
        (75, 58.762)
        (100,58.692)
        (125,58.648)
        (150,58.659)
        (175,59.028)
        (200,58.914)
        (225,58.902)
        (250,58.941)
        (275,59.154)
        (300,58.963)
        (325,59.117)
        (350,58.918)
        (375,59.256)
        (400,59.197)
        (425,59.432)
        (450,59.204)
        (475,59.401)
        (500,59.265)
        };
        \addplot [color=cyan, thin, mark=*, mark size=1pt] plot coordinates {
        (1,  45.395)
        (25, 45.068)
        (50, 44.995)
        (75, 45.109)
        (100,45.01)
        (125,45.188)
        (150,45.03)
        (175,45.203)
        (200,45.216)
        (225,45.181)
        (250,45.243)
        (275,45.931)
        (300,45.198)
        (325,45.449)
        (350,45.236)
        (375,45.442)
        (400,45.232)
        (425,45.577)
        (450,45.422)
        (475,45.967)
        (500,45.443)
        };

        \addplot [color=brightpink, thick, mark=none, mark size=1.2pt] plot coordinates {
        (1,  52.22)
        (25, 46.856)
        (50, 46.695)
        (75, 46.853)
        (100,46.652)
        (125,46.951)
        (150,46.723)
        (175,47.113)
        (200,46.868)
        (225,46.909)
        (250,46.931)
        (275,47.234)
        (300,47.014)
        (325,47.265)
        (350,46.989)
        (375,47.372)
        (400,47.164)
        (425,47.39)
        (450,47.174)
        (475,47.392)
        (500,47.159)
        };
        \addlegendentry{Ref-opt GPU code}
        \addlegendentry{\rotatebox[origin=c]{180}{$\Lsh$} + data-restructure+split-computation}
        \addlegendentry{\rotatebox[origin=c]{180}{$\Lsh$} + shuffle-memory}
        \addlegendentry{\rotatebox[origin=c]{180}{$\Lsh$} + warp-based-execution}
        \addlegendentry{\rotatebox[origin=c]{180}{$\Lsh$} + multiple-computation/thread-block}
        \addlegendentry{\rotatebox[origin=c]{180}{$\Lsh$} + loop-unroll}

    \end{axis}
 \end{tikzpicture}
}
    \caption{Execution time (in ms) for every 25\textsuperscript{th}
    iteration of the SpMV operations with various optimizations on NVIDIA
    GPU.} \label{fig:gpu-optimization} 
\end{figure}

Figure~\ref{fig:gpu-optimization} presents the execution time for different
optimizations we incorporated in an incremental way for every
25\textsuperscript{th} iteration of the SpMV operation.  The benefits of the
data restructuring optimization and effective partitioning of the computations
per thread block are evident in Figure~\ref{fig:gpu-optimization}. We
calculated the average execution time of $500$ iterations of SBBNNLS to compare
performance. We obtained speedups of $2.11\times$ and $1.81\times$ for the
\textit{Naive + data-restructuring + computation-partition} optimization over
the \textit{Ref-opt} GPU code of the \texttt{DSC} and \texttt{WC} operations
respectively. 

In Figure~\ref{fig:ref-code}, the innermost loop is executed sequentially
performing the \texttt{daxpy} operation and the \texttt{dot-product} operation
for the \texttt{DSC} and \texttt{WC} computations respectively. Parallelizing
the innermost loop with minimized synchronization was a major source of
performance improvement for the SpMV operations. We obtained speedups of
$2\times$ and $1.06\times$ for the \texttt{DSC} and \texttt{WC} computations
respectively over the \textit{Naive + data-restructuring +
computation-partition} code by exploiting the fine-grained parallelism
(Section~\ref{subsec:ss}). In addition, we obtained significant speedups of
$2.28\times$ and $1.62\times$ for \texttt{DSC} and \texttt{WC} respectively
when we incorporated the single warp-based thread block optimization
(Section~\ref{subsec:warp}). Furthermore, when each thread block handled
additional computations by allocating multiple atoms, coefficients, or voxels
per thread block (Section~\ref{subsec:multiple}), we obtained speedups of
$1.06\times$ and $1.29\times$ over the single-warp based approach for the
\texttt{DSC} and \texttt{WC} computations respectively. The reason for the
improvement is that we obtained reduced synchronization overheads and
additional data reuse in shared memory for the \texttt{YPtr}
and~\texttt{DPtr}~vectors.

We obtained an additional performance improvement of $8\%$ when we performed
loop unrolling for the \texttt{DSC} operation. However, the same was not
observed for the \texttt{WC} operation. The loop trip count is not statically
known in the case of \texttt{DSC}, and the compiler's heuristic perhaps chose
not to unroll it. However, the innermost loop trip count for \texttt{WC} was
statically known, and our unrolling there did not improve performance.  

Summarizing the results, for the \texttt{DSC} operation, we achieve the best
performance by using the \textit{voxel-based restructuring} and the
\textit{voxel-based computation partitioning} technique, and through a
fine-grained parallelization while utilizing shared memory. For the \texttt{WC}
operation, we achieve the best performance by using the \textit{atom-based
restructuring} and the \textit{coefficient-based partitioning}, and by
extracting fine-grained parallelism using the shuffle instruction.
Additionally, we obtained performance improvements for both the \texttt{DSC}
and \texttt{WC} operations by incorporating GPU-specific optimizations such as
usage of a single warp per thread block and scheduling multiple
computations~per~thread~block.

\begin{table}[!h] 
\caption{Execution time (in minutes) of the SBBNNLS algorithm
    for various tractography algorithms using STN96 dMRI data (with $\bf
    N_\theta=96$).} \label{tab:fascicle}
    \small
    \begin{tabularx}{0.75\linewidth}{X|l|c|c|cccc}
        \toprule
        Fascicles                & Tractography & Voxels & $\Phi$ size & CPU-naive & CPU-opt& Ref-opt& GPU-opt\\ \midrule
        \multirow{5}[1]{*}{50000}& DET          & 151414 & 510.0 MB    & 16.8m     & 1.17m  & 0.555m & 0.146m \\
                                 & PROB         & 162499 & 522.9 MB    & 20.7m     & 1.71m  & 0.972m & 0.157m \\
                                 & iFOD1        & 212874 & 726.7 MB    & 49.7m     & 2.93m  & 1.595m & 0.331m \\
                                 & SD\_STREAM   & 195066 & 497.2 MB    & 12.9m     & 1.13m  & 0.535m & 0.118m \\
                                 & FACT         & 138860 & 372.8 MB    & 7.10m     & 0.68m  & 0.319m & 0.084m \\\midrule
       \multirow{5}[1]{*}{100000}& DET          & 161443 & 688.1 MB    & 30.3m     & 1.76m  & 1.102m & 0.232m \\
                                 & PROB         & 173685 & 692.8 MB    & 40.9m     & 2.16m  & 1.428m & 0.244m \\
                                 & iFOD1        & 231586 & 1.020 GB    & 1h47m     & 5.03m  & 2.722m & 0.557m \\
                                 & SD\_STREAM   & 217742 & 617.9 MB    & 24.3m     & 1.61m  & 0.764m & 0.170m \\
                                 & FACT         & 161120 & 457.2 MB    & 13.2m     & 1.00m  & 0.506m & 0.117m \\\midrule
       \multirow{5}[1]{*}{150000}& DET          & 165843 & 858.8 MB    & 45.8m     & 2.32m  & 1.391m & 0.310m \\
                                 & PROB         & 178984 & 851.6 MB    & 50.1m     & 2.81m  & 1.830m & 0.318m \\
                                 & iFOD1        & 239522 & 1.321 GB    & 2h27m     & 7.53m  & 3.631m & 0.747m \\
                                 & SD\_STREAM   & 227416 & 721.1 MB    & 35.8m     & 2.12m  & 0.930m & 0.216m \\
                                 & FACT         & 171782 & 520.8 MB    & 19.4m     & 1.33m  & 0.641m & 0.130m \\\midrule
       \multirow{5}[1]{*}{200000}& DET          & 168608 & 1.001 GB    & 59.0m     & 2.71m  & 1.644m & 0.387m \\
                                 & PROB         & 182302 & 1006  MB    & 1h19m     & 4.21m  & 2.232m & 0.396m \\
                                 & iFOD1        & 244265 & 1.611 GB    & 3h20m     & 9.27m  & 4.345m & 0.950m \\
                                 & SD\_STREAM   & 233403 & 818.5 MB    & 47.1m     & 2.49m  & 1.124m & 0.262m \\
                                 & FACT         & 178779 & 579.0 MB    & 25.4m     & 1.51m  & 0.720m & 0.156m \\\midrule
       \multirow{5}[1]{*}{250000}& DET          & 170403 & 1.171 GB    & 1h14m     & 3.37m  & 1.852m & 0.459m \\
                                 & PROB         & 184613 & 1.132 GB    & 1h56m     & 4.82m  & 2.616m & 0.471m \\
                                 & iFOD1        & 247356 & 1.905 GB    & 4h09m     & 10.9m  & 5.798m & 1.202m \\
                                 & SD\_STREAM   & 237399 & 915.4 MB    & 58.8m     & 2.94m  & 1.288m & 0.304m \\
                                 & FACT         & 183885 & 633.8 MB    & 31.7m     & 1.83m  & 0.812m & 0.190m \\ \midrule    
       \multirow{5}[1]{*}{500000}& DET          & 175351 & 1.970 GB    & 2h42m     & 5.76m  & 3.039m & 0.829m \\
                                 & PROB         & 190589 & 1.871 GB    & 3h52m     & 8.71m  & 4.485m & 0.859m \\
                                 & iFOD1        & 255309 & 3.362 GB    & 6h05m     & 21.1m  & 9.009m & 2.155m \\
                                 & SD\_STREAM   & 247291 & 888.7 MB    & 1h56m     & 4.85m  & 2.070m & 0.528m \\
                                 & FACT         & 197299 & 1.024 GB    & 1h02m     & 3.08m  & 1.249m & 0.301m \\
        \bottomrule
    \end{tabularx}
\end{table}
\subsection{Analyzing performance by varying various parameters of LiFE}

Table~\ref{tab:fascicle} shows absolute execution time of \textit{CPU-naive},
\textit{CPU-opt}, \textit{Ref-opt} and \textit{GPU-opt} implementations of SpMV
operation used in SBBNNLS for different parameters of the LiFE such as number
of fibers and voxels, and various tractography algorithms on the \textit{DS2}
dataset. As discussed in  Section~\ref{subsec:matrix-computation}, the
\texttt{wPtr} vector becomes sparser as it is updated after every iteration of
SBBNNLS, and also as the number of fascicles and the number of voxels
increases.  Consequently, sparser the vector, higher the number of unnecessary
computations.  Thus, we obtained additional reduction in execution time due to
the sparse property of \texttt{wPtr}. This is evident from
Table~\ref{tab:fascicle} for various tractography algorithms. We also observe
that as the number of voxels increases, the size of the demeaned diffusion
signal vector (\texttt{YPtr}) and the execution time of the SBBNNLS algorithm
also increases. If we consider different tractography algorithms mentioned in
the table for the different number of fascicles, the total time to prune the
connectome takes approximately $44$ hours for \textit{CPU-naive} code version,
and took $2$~hours for the \textit{CPU-opt} code version, that is, an overall
speedup of $22\times$. Similarly, for the GPU implementations, it took
$13.26$~minutes for \textit{GPU-opt} code version, and took $61.2$~minutes for
the \textit{Ref-opt} GPU code version, that is, an overall speedup of
$4.6\times$.

Usually, the LiFE application apart from generating the optimized connectome
for a single tractography algorithm, it also generates optimized connectomes
for various tractography algorithms and the number of fascicles to compare
them. The optimizations we discussed in Section~\ref{sec:optimizations} can be
extended to several tractography algorithms that are used to compute the
optimized connectome. In addition to that, the voxel size for the datasets we
used was $1.5$-$2$~mm; however, if the voxel size is reduced to half, the
memory consumption for a connectome matrix may increase up to $8\times$. For
high-resolution DWI datasets, the voxel size may be as low as
$0.1$~mm~\cite{Stucht2015}, hence the memory utilization for connectome
matrices generated from these datasets can scale to an order of PBs.

\begin{table}[!t]
\begin{center}
    \caption{Execution time (in minutes) up till different iterations of the 
    SBBNNLS for various code implementations running on CPU and GPU.}
    \label{tab:cpu-gpu}
    \small
    \begin{tabularx}{0.84\linewidth}{rXXXXXXXXXX}
        \toprule
        \multirow{2}[1]{*}{Iterarions}     & \multicolumn{5}{c}{Execution time (minutes)}  & \multicolumn{5}{c}{Speedup over}        \\ 
                  \cmidrule(lr){2-6} \cmidrule(lr){7-11}
            & CPU-naive & CPU-opt & Ref-opt & ReAl-LiFE & GPU-opt& CPU-naive& CPU-opt &Ref-opt& ReAl-LiFE &GPU-opt   \\ \midrule
        10  &  5.241    & 0.304   & 0.421   &     0.035 & 0.025  &   1.0    &    17.24& 12.48 &    150.60 & 209.64 \\
        100 &  45.07    & 1.978   & 1.344   &     0.318 & 0.186  &   1.0    &    22.79& 33.54 &    141.41 & 242.35 \\ 
        500 &  224.8    & 8.294   & 4.393   &     1.603 & 0.855  &   1.0    &    27.12& 51.21 &    140.23 & 263.06 \\
        \bottomrule
    \end{tabularx}
\end{center}
\end{table}

\subsection{Comparing execution time in different code implementations} \label{subsec:codesresults} In
Table~\ref{tab:cpu-gpu}, we compare execution time in minutes for various code
implementations of the SpMV operations up till different iterations of SBBNNLS
on CPU and GPU systems. We observe that our \textit{CPU-opt} implementation
achieves an overall speedup of $27.12\times$ over the \textit{CPU-naive}
implementation. Additionally, one can observe that the speedup improves as the
number of iterations increases; the reason for this is due to the
non-negativity constraint (exploited by \texttt{wPtr}) in SBBNNLS.

The speedup that our \textit{GPU-opt} implementation obtains over the
\textit{Ref-opt} implementation is due to the optimizations discussed in
Section~\ref{sec:gpuoptimizations} that helped to obtain better data reuse,
exploit fine-grained parallelization, and minimize synchronization. Whereas,
the speedup we obtain over the \textit{ReAl-LiFE} implementation is due to the
following reasons. 
\begin{enumerate}

\item The \textit{ReAl-LiFE} implementation does not exploit
the sparse property of the \texttt{wPtr} for \texttt{DSC} operation. This can
be seen from Table~\ref{tab:cpu-gpu}, where the speedups for \textit{ReAl-LiFE}
reduce with the different iterations. In contrast, for \textit{GPU-opt}
implementation, the performance improves significantly for different
iterations. Using this property, our \textit{GPU-opt} implementation obtained
an added speedup of $2.51\times$ for average execution of $500$ iterations of
\texttt{DSC}. 

\item \textit{ReAl-LiFE} implementations use the
\textit{voxel-based computation partitioning + voxel-based data restructuring}
combination by default for both the SpMV operations. However, our
implementation achieves the best performance by incorporating the
\textit{voxel-based computation partitioning + voxel-based data restructuring}
combination for \texttt{DSC} (that is similar to \textit{ReAl-LiFE}) and the
\textit{coefficient-based computation partitioning + atom-based data restructuring}
combination for the \texttt{WC}. If we use the combination proposed by
\textit{ReAl-LiFE} then the \textit{GPU-opt} performance drops by $17\%$
over our proposed combination for the SBBNNLS algorithm.  In addition, the best
computation partitioning + data restructuring choice depends on the dMRI
dataset.  Using a fixed choice may result in loss of performance.  Therefore,
our selection is an automatic runtime-based one that dynamically determines the
best partitioning by analyzing the performance of each combination for a dMRI
dataset.  

\item We also schedule multiple computations to a thread block to
enhance data reuse and reduce synchronization (Section~\ref{subsec:multiple}).
This optimization was not incorporated by \textit{ReAl-LiFE}, but when
incorporated for \textit{GPU-opt}, it helped to improve the overall performance
by $1.05\times$ and $1.29\times$ for \texttt{DSC} and \texttt{WC} operations
respectively. 

\item To obtain fine-grained parallelism for the \texttt{WC}
operation, the \textit{ReAl-LiFE} uses shuffle instruction + shared memory,
whereas we used only shuffle instruction. This optimization helped to reduce
the consumption of shared memory; however, in terms of performance, it did not
affect much. 

\item Additionally, the \textit{ReAl-LiFE} approach uses the
syncthread barrier, whereas we used a much cheaper syncwarp operation. Usage of
syncwarp would not help to gain performance for the \textit{ReAl-LiFE}
implementation because it doesn’t incorporate multiple computations per thread
block optimization. On the other hand, if we use syncthread barrier for our
implementation then the performance drops by $10\%$. 
\end{enumerate}

Thus, our approach not only leverages best aspects of both the \textit{Ref-opt} and
the \textit{ReAl-LiFE} implementations, but also complements them by taking
advantage of new optimizations. Hence, our \textit{GPU-opt} implementation
achieves significant speedups of $5.2\times$ and $1.87\times$ over the
\textit{Ref-opt} and \textit{ReAl-LiFE} implementations respectively.

\section{Related Work} \label{sec:relatedwork} In this section, we discuss
prior work on optimizing the compute-intensive sparse matrix vector (SpMV)
operations of the LiFE application. Next, we discuss various approaches
proposed to tackle indirect array accesses and obtain performance improvement
in their presence for CPUs. We also discuss various sparse formats and
optimization techniques proposed to enhance the performance of SpMV for GPUs.

\subsection{Optimizing SpMV operations of the LiFE algorithm} In this section,
we discuss existing implementations to optimize the SpMV operations of  the
LiFE application on various architectures.

\subsubsection{Madhav's GPU Implementation:} \cite{madhav2017} developed
a GPU implementation for the compute-intensive matrix operations of LiFE.
Madhav by default performs the atom-based data restructuring (discussed in
Section~\ref{subsec:ds}) to exploit data reuse and uses the coefficient-based
partitioning (discussed in Section~\ref{subsec:tb}) to achieve
coarse-grained~parallelism. In addition to this, Madhav's GPU implementation
exploits the sparse property of the \texttt{wPtr} vector to avoid unnecessary
operations to further improve the performance. However, the \textit{data
restructuring + computation partitioning} choice used in this implementation
requires an atomic operation to avoid data races (which leads to
synchronization across the thread blocks of a GPU); hence, this results in
significant drop in performance. Our optimized GPU implementation is built upon
it and additionally performs other optimizations discussed in
Section~\ref{sec:gpuoptimizations} to obtain a speedup of $5.2\times$ over it.    

\subsubsection{ReAl-LiFE:} \cite{kumar2019} presented
ReAl-LiFE algorithm, a modification of the LiFE algorithm introducing an
additional regularized constraint to prune connectomes. This work also presents
a GPU implementation of LiFE's SpMV operations. Our GPU implementation obtains
a speedup of $1.87\times$ over the ReAl-LiFE implementation due to the
differences discussed in Section~\ref{subsec:codesresults}.  

\subsubsection{MPI-LiFE} \cite{gugnani17hipc} presented a distributed memory
based design to parallelize the multiplication of large but sparse N-dimension
arrays for the LiFE algorithm. Using the MPI and OpenMP programming models, the
authors used \textit{MPI-based} and \textit{MPI+OpenMP-based} LiFE designs,
collectively named as \textit{MPI-LIFE}, to accelerate the SpMV operations of
the LiFE model. On a single node (KNL-based), the MPI-LiFE model achieved a
speedup of $8.7\times$, and on multiple nodes (16 Intel Xeon SandyBridge-based
ones), a speedup of $8.1\times$, over the original CPU version. The problem of
irregular accesses becomes more prominent with multiple nodes, as the
performance of MPI-LiFE could not scale due to memory latency and bandwidth
bottlenecks. The MPI-LiFE code was not publicly available, and so we could not
evaluate it as a reference.

\subsection{Optimizing irregular applications using insepector/executor
paradigm} Code optimization and transformation frameworks have been studied
well in the literature for improving data locality and parallelism for regular
or affine array
references~\cite{lu1991,feautrier1992a,sarkar1992,li1994,carr1994,kelly1995,
kodukula1996,wolf1996,kandemir1998,cierniak1995,thies2001,pugh1991sc}.  Among
many frameworks, the polyhedral framework is popular for optimization of affine
loop nests~\cite{feautrier1991ijpp,cohen05ics,uday08pldi,verdoolaege2010icms}.
However, most of the literature on the polyhedral framework is inapplicable to
the code with non-affine accesses.

In literature, significant prior work has been proposed to support non-affine
accesses by extending the polyhedral framework \cite{venkat14cgo,venkat15pldi,
venkat2016sc,strout16parallelcomputing}. New representations
\cite{belgin2009ics,bell2009sc,liu2013ics,mellorcrummey2004,vuduc2005a,williams2007sc,
shantharam2011}, transformations~\cite{venkat15pldi,wu2013ppopp,mitchell99pact,
ding1999,hwansooHan2006tpds} and code generation frameworks~\cite{venkat14cgo,
strout16parallelcomputing} have been proposed to achieve the performance
similar to hand-tuned library versions \cite{balay10,bell2009sc,bulu2011ijhpca,
mellorcrummey2004,vuduc2005b}. As discussed earlier, indirect array accesses
cannot be analyzed precisely at compile time. Therefore, most prior work
incorporated an inspector/executor approach to tackle this issue. The inspector
analyzes the code and collects the non-affine access information and executor
uses this information to generate the code. 

\cite{venkat14cgo} based on the inspector/executor
paradigm extended polyhedral code generation to support irregular array
accesses in loop bounds and references. The non-affine accesses were
represented using uninterpreted functions~\cite{pugh1994} and supported loop
coalescing. The work targeted code generation for GPUs involving sparse
matrix-vector multiplication operation and achieved comparable performance to
hand-tuned CUSP library. \cite{venkat15pldi} work
extended \cite{venkat14cgo} by introducing three new compiler transformations
to represent and transform sparse matrix computations. The work generated
optimized code for the sparse representations and targeted reduction in runtime
overhead. Both the works were restricted to non-affine read-only accesses for
sparse matrix computations.  Whereas, our approach uses an custom approach to
obtain data reuse and is able to handle multiple read and write non-affine
array accesses with a much lower overhead than the proposed works. Our approach
is specialized and can be used for STD-based sparse matrix operations and
representations. However, targeting optimization of different sparse
representation is not the target of this paper and can be future work.

Furthermore, in another work presented by \cite{venkat2016sc} demonstrates parallelized code generation for sparse
matrix applications such as ILU factorization and Gauss-Seidel relaxation,
having loop-carried dependences. The proposed work is specialized to
automatically generate the runtime inspector and executor to achieve wavefront
parallelization; exploiting fine-grained parallelism by parallelizing within
the wavefront and synchronizing (by using OpenMP barriers) across the
wavefronts, hence, introducing pipelined-startup stalls and synchronization
overhead across the wavefronts. However, our work to parallelize the sparse
code is specialized to specific structure and sparsity of matrices used in the
LIFE algorithm that not only exploits coarse-grained parallelism (marking
outermost-loop parallel using OpenMP) without synchronization but also utilizes
the fine-grained parallelism (utilizing vectorization by usage of a BLAS call). 

\cite{strout16parallelcomputing} develops a {"sparse
polyhedral framework"} (SPF), a code generation approach to utilize data
locality in applications involving non-affine array index and loop bounds. SPF
specifies runtime reordering transformations and algorithms to automatically
generate inspector/executor code to implement these transformations. The
generated code competes with hand-optimized ones but requires additional time
for representation, inspection, transformation and executor code generation.
The time required by an inspector is amortized over different iterations of the
program. However, our inspector approach utilizes both data locality and
parallelism, though, limited to single level indirect array access (i.e.
A[B[i]]). In addition, our approach presents a specific inspector model
utilizing data reordering transformation and doesn't require an additional
overhead of code generation. Moreover, our approach significantly reduces the
time required by the inspector by amortizing it over different runs of the
program as seen in the SBBNNLS algorithm of the LiFE algorithm.

\subsection{Optimizing SpMV operations for GPUs} SpMV is a widely used kernel
operation for a large number of applications. A number of sparse
representations~\cite{Ekambaram2003iscis,Benatia2016ipcc,Yang2018ipcc,Mahmoud2017csci}
have been proposed to avoid unnecessary computations and tackle the memory
bottleneck. Based on the sparse representation technique used, the memory
accesses may vary from moderately regular to highly irregular ones, posing a
challenging problem.  Exploiting the massive parallelism and multi-threaded
processing power of architectures such as GPUs makes the challenge even more
tougher due to the load imbalance issue and a different multi-level memory
hierarchy when compared to CPUs. Many prior works  introduced new storage
formats~\cite{liu2013ics,belgin2009ics,bell2009sc} and various optimization
techniques~\cite{mellorcrummey2004,Vzquez2010,Greathouse2014ichpc,Choi2010ppopp}
to address this challenge.

One of the earliest works to optimize SpMV kernel for GPUs was of \cite{Baskaran2009OptimizingSM}. They addressed two key aspects
involved in optimizing SpMV for GPUs: thread mapping and data access strategies
for compressed sparse row (CSR) format. They presented various optimization
techniques such as exploiting synchronization-free parallelism, optimized
thread-mapping, and optimized off-chip memory access to improve performance of
SpMV. In another work to optimize SpMV, \cite{bell2009sc}
incorporated specific optimization techniques to exploit regularity patterns
for different sparse representation techniques such as DIA, ELL, COO and CSR
formats. Further, they presented a new sparse matrix representation named ---
``Hybrid'', to improve the performance of SpMV. 

Prior works on optimizing SpMV have focused on techniques tailored for a
specific sparse representation to exploit structure in irregular accesses.
However, there are a large class of problems involving large matrices that are
better solved using a tensor decomposition approach to reduce memory
requirements. Low-rank Sparse Tucker Decomposition (STD) is one such popular
tensor decomposition technique used for numerous applications performing matrix
operations. The sparse representations may involve multiple indirect array
accesses, making the problem hard; however, this is a necessary trade-off
considering the reduction obtained in memory requirement.

Other works on optimizing GPU applications performing SpMV operations using the
Tucker decomposition have focused on the dense matrix
operations~\cite{ChoiLC18sc,Chakaravarthy2018ics}, or a distributed memory
system based STD approach targeting tensor-times-matrix
operation~\cite{Kaya2016icpp,Chakaravarthy2018ics,Choi2018ipdps,Perros2015icdm}.
In contrast, we proposed several optimization techniques for the STD-based SpMV
operations used in LiFE. Our data restructuring and computation partitioning
optimizations could potentially be generalized and extended to other
applications employing STD, although one would have to look for similar or
other data patterns.  Furthermore, other alternatives to STD such as Kronecker
Product and CANDECOMP/PARAFAC methods could also potentially benefit from our
optimizations. 
 
\section{Conclusions} \label{sec:conclusions} 
We addressed challenges involved in optimizing the SpMV operations for large
matrices in conjunction with a popular tensor decomposition technique, namely,
Sparse Tucker Decomposition (STD). The matrices when represented using the STD
technique involved several indirect accesses and exhibited poor performance.
LiFE algorithm is a popular neuroscience application in which large-sparse
matrices are represented using STD. Once these matrices were decomposed to a
sparse-tensor format, the SpMV operations of LiFE were transformed into a complex
sequence of operations, involving multiple indirect accesses. 

First of all, we proposed target-independent optimization techniques to
optimize matrix operations of LiFE such as: (1)~standard compiler optimizations
to avoid redundant computations, (2)~a custom data restructuring
technique to exploit data reuse and minimize the downsides of irregular
accesses; this optimization in turn made other optimizations valid and
fruitful, and (3)~methods to partition computation among threads to exploit
coarse-grained parallelism while reducing synchronization overhead. Then we
presented target-specific optimizations for CPU and GPU systems. The
CPU-specific optimizations that we incorporated includes efficient
synchronization-free thread scheduling and mapping appropriate code fragments
to a BLAS call in the SpMV operations. Our highly optimized parallel CPU
implementation utilized the target-independent optimizations and tailored these
CPU-specific optimizations for LiFE application to obtain a speedup of
$27.12\times$ over the original sequential CPU approach (running on 16 core
Intel Xeon Silver system). We also extend the PolyMage DSL to automatically
generate an optimized CPU code for the SpMV operations of the LiFE as a
proof-of-concept.  Next, we presented GPU-specific optimizations such as:
(1)~exploiting fine-grained parallelism by utilizing shared memory and the
shuffle instruction, (2)~map multiple computations to a single thread block to
exploit additional data reuse, and (3)~transform loops to minimize
synchronization. We utilized target-independent optimizations and tailored
these GPU-specific optimizations to optimize the SpMV operations of the LiFE
application, which when executed on an NVIDIA's GeForce RTX 2080 Ti GPU,
achieved speedups of $5.2\times$ and $1.87\times$ respectively over an existing
optimized GPU implementation and over the ReAl-LiFE implementation. In the
future, we plan to extend our work to support other STD-based applications, and
to also design domain-specific abstractions and code generation support in
existing frameworks to automate these tasks.

\begin{acks}
  We are deeply grateful to Dr. Sridharan Devarajan and Varsha Sreenivasan from 
  the Centre for NeuroScience, Indian Institute of Science for introducing us to 
  the neuroscience domain context associated with this work, and for help with 
  writing the introduction and background sections of this paper.  This work was 
  supported in part by a grant (EMR/2016/008015) from the Science and 
  Engineering Research Board (SERB), India through its Extramural Research 
  funding program.
\end{acks}

\bibliographystyle{ACM-Reference-Format}
\bibliography{bibfile}

\end{document}